\documentclass[12pt]{article}
\usepackage[utf8]{inputenc}
\usepackage[T1]{fontenc}
\usepackage{amsmath,amssymb,amsthm,commath,enumitem,mathtools}
\usepackage{bm}
\usepackage{kbordermatrix}
\usepackage[singlelinecheck=false]{caption}
\usepackage[margin=2.54cm]{geometry}
\usepackage[normalem]{ulem}
\usepackage{natbib}
\usepackage{graphicx}
\usepackage{booktabs}
\usepackage{color} 
\usepackage{url} 
\usepackage[onehalfspacing]{setspace}
\usepackage[small]{titlesec}
\usepackage{enumitem}
\usepackage[ruled,vlined]{algorithm2e}
\usepackage[tableposition=top,labelfont=bf]{caption}
\usepackage{layouts}
\usepackage{xcolor}
\usepackage{makecell}
\graphicspath{ {./figures/} }
\usepackage{fancyhdr}

\allowdisplaybreaks
\let\originalleft\left
\let\originalright\right
\renewcommand{\left}{\mathopen{}\mathclose\bgroup\originalleft}
\renewcommand{\right}{\aftergroup\egroup\originalright}
\newcommand{\overbar}[1]{\mkern 1.5mu\overline{\mkern-1.5mu#1\mkern-1.5mu}\mkern 1.5mu}

\title{%
   Zero-state Coupled Markov Switching Count Models for Spatio-temporal Infectious Disease Spread \\[5pt]
  }
\author{Dirk Douwes-Schultz\footnote{{{\it Corresponding author}: Dirk Douwes-Schultz, Department of Epidemiology, Biostatistics and Occupational Health, McGill University, 2001 McGill College Avenue, Suite 1200, Montreal, QC, Canada, H3A 1G1. {\it E-mail}: {\tt
				dirk.douwes-schultz@mail.mcgill.ca}.}} \hspace{1mm} and Alexandra M. Schmidt \\
				\textit{Department of Epidemiology, Biostatistics and Occupational Health} \\ \textit{McGill University, Canada }}
				
\date{\today}

\begin{document}

\maketitle

\begin{abstract}
Spatio-temporal counts of infectious disease cases often contain an excess of zeros. With existing zero inflated count models applied to such data it is difficult to quantify space-time heterogeneity in the effects of disease spread between areas. Also, existing methods do not allow for separate dynamics to affect the reemergence and persistence of the disease. As an alternative, we develop a new zero-state coupled Markov switching negative binomial model, under which the disease switches between periods of presence and absence in each area through a series of partially hidden nonhomogeneous Markov chains coupled between neighboring locations. When the disease is present, an autoregressive negative binomial model generates the cases with a possible 0 representing the disease being undetected. Bayesian inference and prediction is illustrated using spatio-temporal counts of dengue fever cases in Rio de Janeiro, Brazil.

{\bf Key words :} Bayesian paradigm; Dengue fever; Forward filtering backward sampling; Zero-inflation; Hidden Markov model; State space model. \end{abstract}

\section{Introduction}

In epidemiology, counts of infectious disease cases are being increasingly reported in several related areal units across time. A common issue encountered when modeling these counts is the presence of excess zeros \citep{arabSpatialSpatioTemporalModels2015}. That is, there are often many more zeros in the counts than can be predicted by the usual Poisson and negative binomial count models. \cite{lambertZeroInflatedPoissonRegression1992} proposed the zero inflated Poisson (ZIP) model to deal with excess zeros in count data, and this approach, with various extensions (see \cite{youngZeroinflatedModelingPart2020}), has been applied in many fields including epidemiology. We will refer to any model that incorporates zero-inflation into a count distribution as a zero inflated count (ZIC) model (e.g. ZIP, ZINB, ZIGP, ZICMP models, etc. \citep{youngZeroinflatedModelingPart2020}). In a disease mapping application of the ZIC model, the presence/absence of the disease is generated through a Bernoulli process and then, when the disease is present, the number of reported cases is generated through a count process, typically negative binomial or Poisson \citep{fernandesModellingZeroinflatedSpatiotemporal2009}. A zero coming from the count process represents the disease being undetected, while a zero from the Bernoulli process represents the true absence of the disease. A ZIC model can help account for the excess zeros, through the additional Bernoulli zero generation, and also allows us to investigate factors related to the presence/absence of the disease while considering imperfect detection of disease presence \citep{vergneModellingAfricanSwine2016}. ZIC models have become increasingly popular in epidemiology for modeling excess zeros in spatio-temporal infectious disease counts \citep{fernandesModellingZeroinflatedSpatiotemporal2009,aktekinAnalysisIncomeInequality2015,wangdiSpatialTemporalPatterns2018}.

However, in a spatio-temporal setting, an important issue with existing ZIC approaches is that they only model the probability of overall disease presence. In a spatio-temporal setting we can separate the presence of a disease into two events of epidemiological interest, persistence (presence to presence) and reemergence (absence to presence). Some important epidemiological covariates can affect the reemergence and persistence of an infectious disease quite differently. An example  from our application of dengue fever is temperature, as temperature affects both the hatching rate of vertically infected mosquito eggs, which is more important for the reemergence of the disease, and the life cycle of the infected mosquitoes, which is more important for the persistence of the disease \citep{coutinhoThresholdConditionsNonAutonomous2006}. 

A second issue concerns the use of random effects in the Bernoulli process of a ZIC model, which can be used to account for spatio-temporal correlations in the presence of the disease \citep{hoefSpaceTimeZeroinflated2007,torabiZeroinflatedSpatiotemporalModels2017,giorgiBivariateGeostatisticalModelling2018}. Splines can also be used for this purpose \citep{ghosalHierarchicalMixedEffect2020}, but are functionally similar so we will refer in general to random effect models. Random effect models are built by specifying relationships between probabilities of disease presence between areas and across time \citep{torabiZeroinflatedSpatiotemporalModels2017}. In contrast, it is more traditional in epidemiology to build models of infectious disease spread by specifying the probability that the disease will be present in an area given its actual presence in the area and neighboring areas previously \citep{keelingDynamics2001UK2001,smithPredictingSpatialDynamics2002,hootenStatisticalAgentBasedModels2010}. {\color{black}That is, random effects models condition on the probability of disease presence in neighboring areas while epidemiological models usually condition on the actual presence of the disease in neighboring areas.} If a goal of the analysis is to quantify associations between various space-time factors and the effects of disease spread between areas, like in \cite{smithPredictingSpatialDynamics2002} who examined whether rivers block the spread of rabies, then the traditional epidemiological approach is more appropriate. This is because the disease can only spread from an area if it is present there, {\color{black} while the probability of disease presence does not tell you if the disease is spreading or not. For example, the disease could be present in a neighboring area and spread from there even if its probability of presence is low.} Therefore, it is difficult to quantify space-time heterogeneity in the effects of disease spread between areas with existing ZIC models.

To help address these issues we propose a zero-state Markov switching approach to modeling the zero inflation. Markov switching models, first introduced by \cite{goldfeldMarkovModelSwitching1973} and extensively developed by \cite{hamiltonNewApproachEconomic1989}, allow a time series to be described by several submodels (states) where switching between submodels is governed by a first order Markov chain. The motivation behind the Markov switching model, over a finite mixture model, is that the states are often dependent across time and occur consecutively \citep{goldfeldMarkovModelSwitching1973}. Markov switching models that switch between a zero state and a count state were considered in a temporal setting by \cite{wangMarkovZeroinflatedPoisson2001} and in a spatio-temporal setting by \cite{malyshkinaZerostateMarkovSwitching2010}. However, \cite{malyshkinaZerostateMarkovSwitching2010} assumed the states were independent between spatial units and neither of these applications dealt with infectious disease data. 

In our framework, following traditional disease mapping {\color{black}ZIC} models \citep{fernandesModellingZeroinflatedSpatiotemporal2009}, we assume the existence of a zero state, representing the absence of the disease, and a {\color{black}negative binomial} state, representing the presence of the disease. Then we allow the disease to switch between the presence and absence state in each area through a nonhomogeneous \citep{dieboldRegimeSwitchingTimevarying1993} Markov chain. {\color{black} The Markov chains allow for switching between long periods of disease presence and absence, and allow for each covariate to have a separate effect on the reemergence of the disease compared to the persistence.} As the disease can spread between areas, we extend the zero-state Markov switching count model to a coupled Markov switching model \citep{pohlePrimerCoupledStateswitching2020c}, by coupling the Markov chains between neighboring areas. In a coupled Markov switching model, the transition probabilities of a Markov chain can depend on the states of other Markov chains. {\color{black} We use a collection of coupling parameters to account for heterogeneity in the effects of neighboring disease spread across space and time. The coupling parameters allow us to quantify associations between the effects of neighboring disease spread and various space-time factors related to either area.}

This paper is structured as follows. In Section 1.1 we introduce our motivating example of dengue fever cases in Rio de Janeiro and lay out the goals for our analysis. In Section 2 we introduce our proposed model, the zero-state coupled Markov switching {\color{black}negative binomial} model. In Section 3 we describe Bayesian inference using data augmentation through Markov chain Monte Carlo (MCMC) methods. Here we discuss efficient sampling strategies for the parameters and unknown state indicators. In Section 4 we apply the model to the Rio's dengue fever data. Finally, we close with a general discussion in Section 5.

\subsection{Motivating example: dengue fever in Rio de Janeiro}

\begin{figure}[!hbt]
 	\centering
 	\includegraphics[width=.75\textwidth]{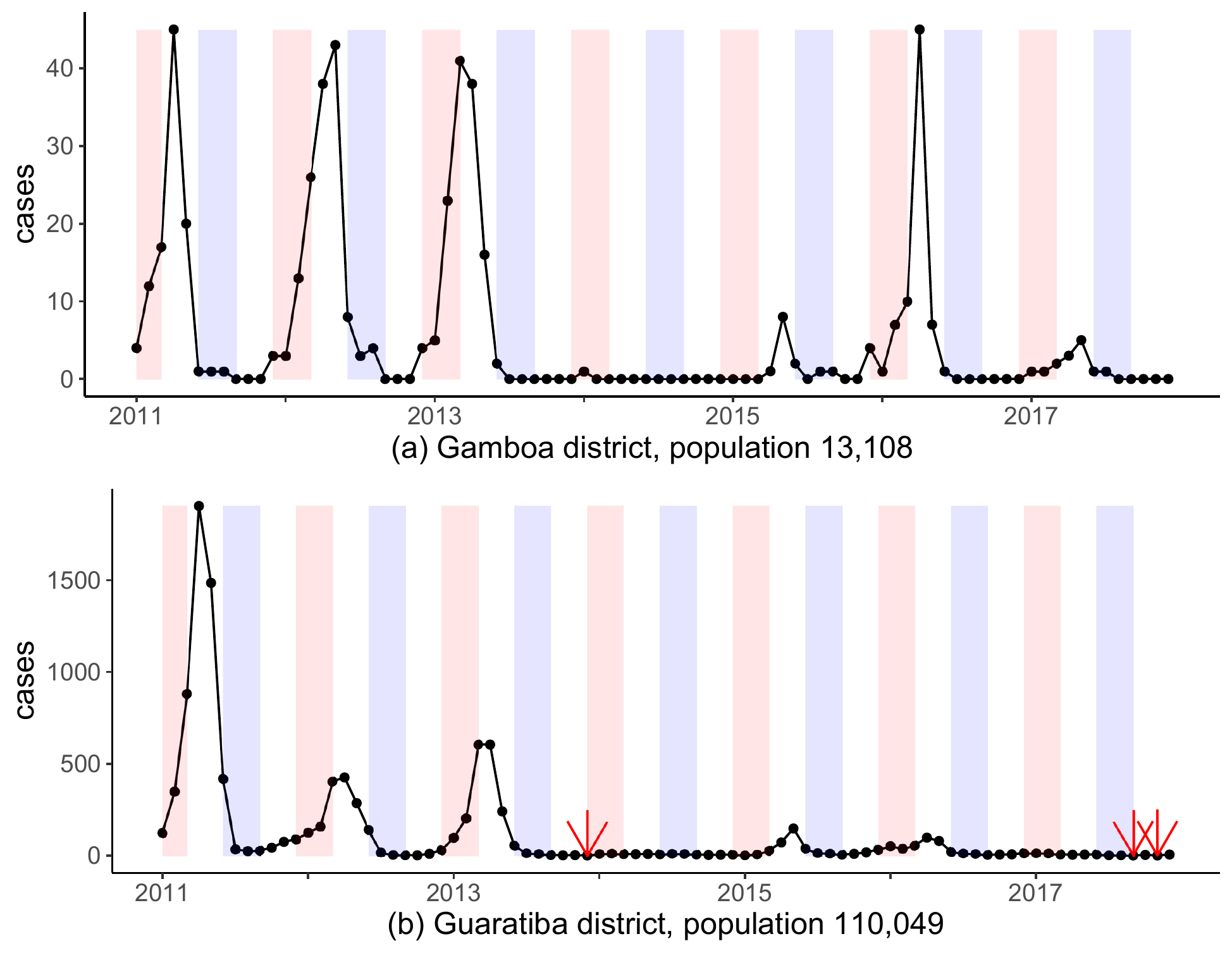}
	\caption{Monthly cases of dengue fever for 2 districts. Summer/winter seasons highlighted in red/blue. In (b), the 3 red arrows point to the only 3 months with 0 cases in this district during the study period. \label{fig:month_cases}} 
\end{figure}

\begin{figure}[ht]
 	\centering
 	\includegraphics[width=.8\textwidth]{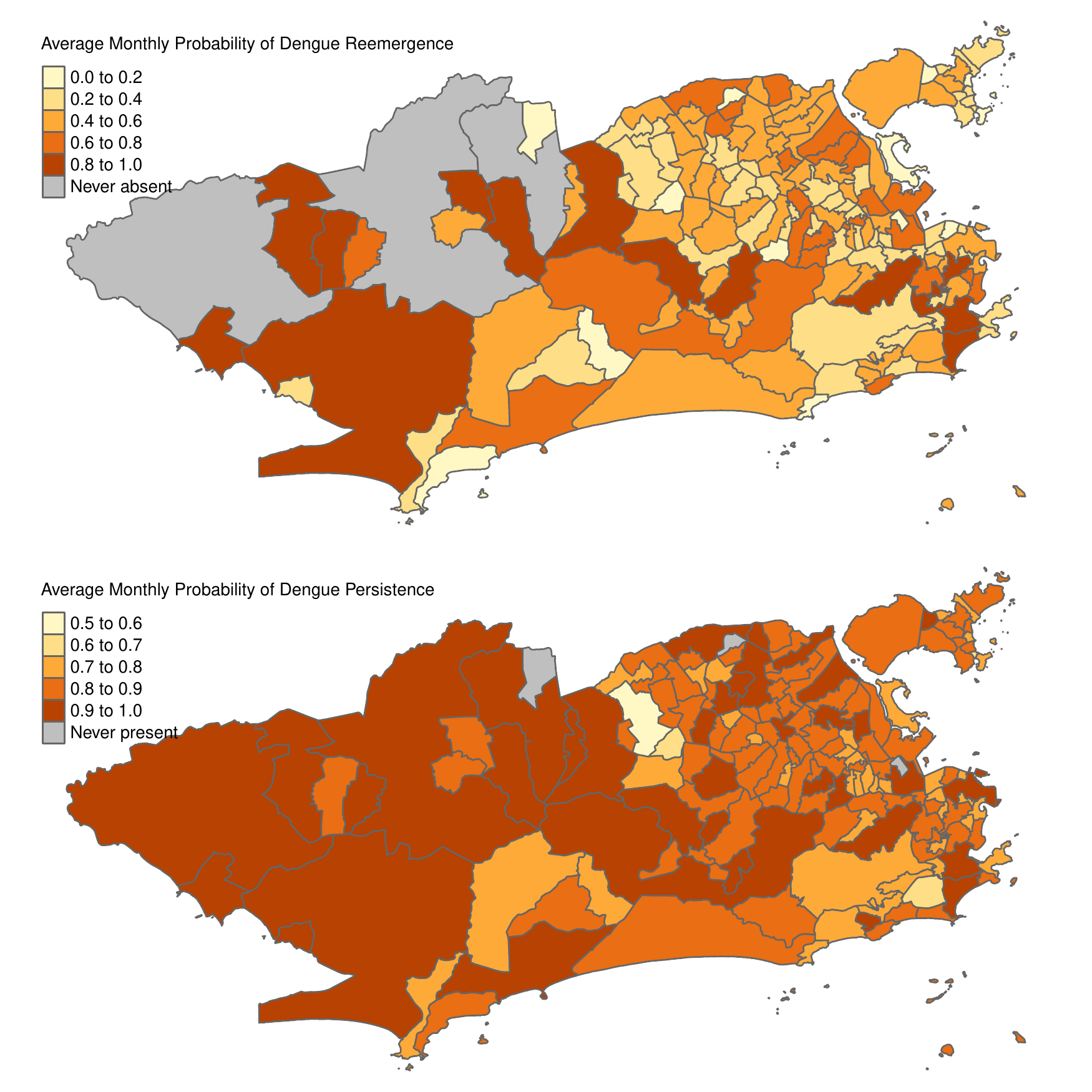}
	\caption{(top) Average monthly probability of dengue reemergence based on reported cases, i.e. $\frac{\#(\text{cases}=0\to \text{cases}>0)}{\#(\text{cases}=0\to \text{cases}>0)+\#(\text{cases}=0\to \text{cases}=0)}$. (bottom) Average monthly probability of dengue persistence based on reported cases, i.e. $\frac{\#(\text{cases}>0\to \text{cases}>0)}{\#(\text{cases}>0\to \text{cases}>0)+\#(\text{cases}>0\to \text{cases}=0)}$. \label{fig:avg_time}} 
\end{figure}

Dengue fever is endemic in Rio de Janeiro, Brazil, and there have been several major epidemics there since 1987. We focus on modeling monthly cases of dengue fever in the 160 districts of Rio de Janeiro between 2011-2017, as reported by the Health Secretary for the city (\url{http://www.rio.rj.gov.br/web/sms/exibeconteudo?id=2815389}).  Figure \ref{fig:month_cases} shows monthly dengue fever cases for two districts, a relatively small district (a) and relatively large district (b). In the smaller district, there are both long periods of dengue presence (9 months longest period) and long periods of dengue absence (14 months longest period). Additionally, in the smaller district there is clearly a recurring seasonal pattern to the persistence and reemergence of the disease. The disease has a lower chance of persisting during the winter, where it often goes extinct, and then dengue often reemerges in the summer. This could be due to changes in rainfall and temperature, which often lead to large fluctuations in the mosquito population \citep{marquardtBiologyDiseaseVectors2004}. Additionally, temperature and rainfall play an important role in the life cycle of vertically infected eggs, which are important for the reemergence of the disease \citep{coutinhoThresholdConditionsNonAutonomous2006}. The same patterns in disease presence are not seen in the larger district {\color{black}in Figure \ref{fig:month_cases} (bottom)}, where dengue persists for much longer on average compared to the smaller district, rarely goes extinct and reemerges quickly. Figure \ref{fig:avg_time} similarly shows that there is significant spatial variation in the district level average, across time, probabilities of dengue reemergence and persistence (although these are based on reported cases and do not account for imperfect detection of disease presence). This could be explained by differences in population between districts \citep{bartlettMeaslesPeriodicityCommunity1957}, but also possibly differences in socioeconomic factors as mosquito's often use human made water containers to lay eggs \citep{schmidtPopulationDensityWater2011}.

A statistical model can help quantify some of the above patterns. Specifically, the goal of our analysis is to investigate how certain risk factors (such as rainfall, temperature, socioeconomic factors and population size) are related to the reemergence and persistence of the disease{\color{black}, and if any effects are different for the reemergence compared to the persistence.} Additionally, as \cite{stoddardHousetohouseHumanMovement2013} showed that many individuals are infected by dengue outside their homes, we want to quantify the risk of dengue spreading between neighboring districts. {\color{black}We want to investigate whether certain space-time factors are associated with an increased risk of dengue spreading from a neighboring district. The number of reported cases in the neighboring district should be important in determining the risk of spread, but also, individuals are known to move more through high population areas \citep{grenfellTravellingWavesSpatial2001}.} Finally, we wish to design some useful warning/forecasting systems for policy makers. 

\section{A zero-state coupled Markov switching negative binomial model}

Assume we have areal data with $i=1,...,N$ areas and $t=1,...,T$ time periods. Let $y_{it}$ be the reported case count from area $i$ during time $t$. Let $S_{it}$ be a binary indicator for the true presence of the disease, meaning $S_{it}=1$ if the disease is present in area $i$ during time $t$ and $S_{it}=0$ if the disease is absent. Finally, let $\bm{S_{(-i)(t-1)}}=\{S_{j(t-1)}\}_{j \neq i}$ be the set containing the state indicators of all areas excluding area $i$ at time $t-1$; and let $\bm{y^{(t-1)}}=(\bm{y_1},...,\bm{y_{t-1}})^T$ be the vector of all case counts up to time $t-1$, where $\bm{y_t}=(y_{1t},...,y_{Nt})^T$. 

We assume that when the disease is present the reported cases are generated by a {\color{black}negative binomial} distribution and that when the disease is absent no cases will be reported, that is, 
\begin{align}
&y_{it} \mid S_{it},\bm{y^{(t-1)}}  \sim \begin{cases} 
        0, & \text{if $S_{it} =0$}  \\[5pt]
    {\color{black} NB(\lambda_{it},r_{it})}, &  \text{if $S_{it} =1$},
   \end{cases} \label{eqn:y_spec} 
\end{align} where $\lambda_{it}$ is the expected number of reported cases given the disease is present {\color{black}and $r_{it}$ is an overdispersion parameter such that $Var\left(y_{it} \mid   S_{it}=1,\bm{y^{(t-1)}}\right)=\lambda_{it}\left(1+\lambda_{it}/r_{it}\right)$}. {\color{black} Zero reported cases may arise when the disease is present due to the disease being undetected by the surveillance system. Therefore, when $y_{it}=0$ we do not observe $S_{it}$ as the disease could be absent or present but undetected.} In general we assume the following forms for $\lambda_{it}$ {\color{black}and $r_{it}$},
 \begin{align}
\lambda_{it}=g\left({\color{black}\bm{\delta}},\bm{x_{it}},\bm{y^{(t-1)}}\right) {\color{black}\text{ and }r_{it}=h\left(\bm{\gamma},\bm{w_{it}},\bm{y^{(t-1)}}\right)}, \label{eqn:lambda}
\end{align} 
where $g(\cdot)$ {\color{black}and $h(\cdot)$} are positive valued functions, {\color{black}$\bm{\delta}$} {\color{black}and $\bm{\gamma}$} are vectors of unknown parameters (including possibly random effects), and $\bm{x_{it}}$ {\color{black}and $\bm{w_{it}}$} are vectors of covariates. Therefore, $\lambda_{it}$ {\color{black} and $r_{it}$} can depend on past values of the case counts, to possibly account for temporal autocorrelation due to transmission of the disease, but not on past values of the state indicators, which is a common assumption in Markov switching models that greatly simplifies model fitting \citep{fruhwirth-schnatterFiniteMixtureMarkov2006}.

To model the switching between periods of disease presence and absence in area $i$, we assume that $S_{it}$ follows a two state nonhomogeneous Markov chain conditional on $\bm{S_{(-i)(t-1)}}$ and $\bm{y^{(t-1)}}$. We propose the following conditional transition matrix for the Markov chain, for $t=1,...,T$,
\begin{align}
&\Gamma\left(S_{it}|\bm{S_{(-i)(t-1)}},\bm{y^{(t-1)}}\right) =
\kbordermatrix{
\textbf{State}  & S_{it}=0 \, \textbf{(absence)} & &  S_{it}=1 \, \textbf{(presence)} \\[8pt]
  S_{i(t-1)}=0 \, \textbf{(absence)} & 1-p01_{it}& & p01_{it} \\[5pt]
  S_{i(t-1)}=1 \, \textbf{(presence)} & 1-p11_{it} & & p11_{it}
  }, \label{eqn:tm}
\end{align} where,
\begin{align*}
&p01_{it} = P\left(S_{it}=1|S_{i(t-1)}=0,\bm{S_{(-i)(t-1)}},\bm{y^{(t-1)}}\right) \qquad \text{(probability of disease reemergence),} \\
&p11_{it} = P\left(S_{it}=1|S_{i(t-1)}=1,\bm{S_{(-i)(t-1)}},\bm{y^{(t-1)}}\right) \qquad \text{(probability of disease persistence)}.
\end{align*} The probability of disease reemergence in area $i$ during time $t$, $p01_{it}$, is allowed to depend on a K-dimensional vector of space-time covariates $\bm{z_{it}}=(z_{it1},...,z_{itK})^T$ {\color{black} as well as the presence of the disease in neighboring areas during the previous time period},
\begin{align}
&\text{logit}(p01_{it}) = \zeta_0 + \bm{z_{it}^T} \bm{\zeta} +\sum_{j \in NE(i)} {\color{black}\phi_{01,j \to i}^{(t-1) \to t}} \, S_{j(t-1)}, \label{eqn:reemergence} 
\end{align} where $NE(i)$ is the set of all neighboring locations of location $i$ {\color{black}and $\phi_{01,j \to i}^{(t-1) \to t}$ is a coupling parameter that represents the effect that the disease spreading from neighboring area $j$ has on the reemergence of the disease in area $i$ at time $t$, the specification of which we will give shortly}. The probability of disease persistence, $p11_{it}$, is defined similarly but all parameters can differ,
\begin{align}
&\text{logit}(p11_{it}) = \eta_0 + \bm{z_{it}^T} \bm{\eta} + \sum_{j \in NE(i)} {\color{black}\phi_{11,j \to i}^{(t-1) \to t}} \, S_{j(t-1)}. \label{eqn:persistence}
\end{align} 
The transition probabilities can depend on $\bm{y^{(t-1)}}$ through $\bm{z_{it}}$, which may contain transformed lagged values of the reported cases (e.g. $\log (y_{i(t-1)}+1)$). This can be justified by the fact that the disease will only go extinct when there is a small number of infectious individuals.  We also need to specify an initial state distribution for the Markov chain in each area, i.e. $p(S_{i0})$ for $i=1,...,N$. 

{\color{black}In order to account for space-time heterogeneity in the effects of neighboring disease spread and to quantify associations between the effects of neighboring disease spread and various space-time factors we use linear models for the coupling parameters \citep{smithPredictingSpatialDynamics2002}, 

\begin{align}
\begin{split}
&\phi_{01,j \to i}^{(t-1) \to t} = \zeta_{0}^{(c)}+\bm{z_{01,ijt}^{(c)T}}\bm{\zeta^{(c)}} \\
&\phi_{11,j \to i}^{(t-1) \to t} = \eta_{0}^{(c)}+\bm{z_{11,ijt}^{(c)T}}\bm{\eta^{(c)}},
\end{split} \label{eqn:coupled_params}
\end{align} where $\bm{z_{01,ijt}^{(c)T}}$ and $\bm{z_{11,ijt}^{(c)T}}$ are vectors of space-time covariates, possibly different, related to either area, e.g. $z_{01,ijt,k}^{(c)}=\log(y_{j(t-1)}/pop_j+1)$ (reported prevalence in the neighboring area) or $z_{01,ijt,k}^{(c)}=\log(pop_i \times pop_j)$ (a gravity term \citep{tuiteCholeraEpidemicHaiti2011}). This is justified as effects of disease spread between areas often vary across space and time with observed factors, some examples being: rabies is less likely to spread between areas separated by rivers \citep{smithPredictingSpatialDynamics2002}; hand, foot, and mouth disease is more likely to spread between farms with many cattle \citep{keelingDynamics2001UK2001}; and measles is more likely to spread from large cities \citep{grenfellTravellingWavesSpatial2001}. We allow the effect of neighboring disease spread on the reemergence of the disease to be different from its effect on the persistence of the disease as these represent two distinct epidemiological processes. The disease may spread to non-infected areas from neighboring areas (modeled through $\phi_{01,j \to i}^{(t-1) \to t}$ in (\ref{eqn:reemergence})); but also, neighboring areas can supply infectious individuals needed to maintain the persistence of the disease in infected areas (modeled through $\phi_{11,j \to i}^{(t-1) \to t}$ in (\ref{eqn:persistence})) \citep{okanoHIVTransmissionSource2020}.}

{\color{black}As the coupling terms in (\ref{eqn:reemergence})-(\ref{eqn:persistence}) account for disease spread from neighboring areas, the $\zeta_0 + \bm{z_{it}^T} \bm{\zeta}$ and $\eta_0 + \bm{z_{it}^T} \bm{\eta}$ terms account for long distance dispersal of the disease as well as within area disease dynamics, i.e. spread from human and/or vector populations within the area.} We assume the effect of covariate $z_{itk}$ on the reemergence of the disease, represented by $\zeta_k$ in (\ref{eqn:reemergence}), can be different from its effect on the persistence of the disease, represented by $\eta_k$ in (\ref{eqn:persistence}). The advantage of considering differing effects for each covariate on the reemergence and persistence of the disease is that many important epidemiological covariates can interact with the reemergence and persistence of an infectious disease quite differently, like temperature in our motivating example.

{\color{black}A final important motivation of the Markov chain model (\ref{eqn:tm}) for $S_{it}$ is that it can account for both many consecutive periods of disease presence and absence. The probability the disease will be absent in location $i$ during time $t$ given it was absent at time $t-1$ is given by $1-p01_{it}$, that is, $1-p01_{it}$ is the probability of having a consecutive period of disease absence. Therefore, as the probability of disease reemergence, $p01_{it}$, approaches 0, consecutive periods of disease absence become more likely. Similarly, as the probability of disease persistence, $p11_{it}$, approaches 1, consecutive periods of disease presence become more likely. Therefore, {\color{black}when} $p01_{it}<<p11_{it}$ the Markov chain can model switching between long periods of disease absence and long periods of disease presence which is often observed for infectious diseases in smaller areas, see Figure \ref{fig:month_cases} (a) and \cite{adamsHowImportantVertical2010}.} 

{\color{black}We will refer to the model defined by (\ref{eqn:y_spec})-(\ref{eqn:coupled_params}) as the zero-state coupled Markov switching negative binomial (ZS-CMSNB) model. {\color{black}Finally, we note that a Poisson version of the ZS-CMSNB model is a special case as $r_{it}$ goes to infinity, which we will refer to as a zero-state coupled Markov switching Poisson (ZS-CMSP) model.}}

\section{Inferential Procedure}
Let $\bm{S_{t}}=\left(S_{1t},...,S_{Nt}\right)^T$ be the set of all state indicators at time $t$ and let $\bm{S}=\{\bm{S_{t}}\}_{t=0}^{T}$ be the set of all state indicators. {\color{black} The vector} $\bm{S_{t}}$ forms a first order Markov chain, conditional on $\bm{y^{(t-1)}}$, with state space $\{0,1\}^N$ and $2^{N}X2^{N}$ transition matrix $\Gamma\left(\bm{S_{t}}|\bm{y^{(t-1)}}\right)$. An element of $\Gamma\left(\bm{S_{t}}|\bm{y^{(t-1)}}\right)$ is given by, 
\begin{align}
\begin{split}
& P\left(S_{1t}=s_{1t},...,S_{Nt}=s_{Nt}|S_{1(t-1)}=s_{1(t-1)},...,S_{N(t-1)}=s_{N(t-1)},\bm{y^{(t-1)}}\right) \\
& \qquad = \prod_{i=1}^{N} P\left(S_{it}=s_{it}|\bm{S_{t-1}}=\bm{s_{t-1}},\bm{y^{(t-1)}}\right). \label{UM}
\end{split}
\end{align} Therefore, rewriting the {\color{black}ZS-CMSNB model} in terms of $\Gamma(\bm{S_{t}}|\bm{y^{(t-1)}})$ and $p(\bm{y_t}|\bm{S_{t}},\bm{y^{(t-1)}})=\prod_{i=1}^{N} p(y_{it}|S_{it},\bm{y^{(t-1)}})$ shows that it is a Markov switching model as defined by \cite{fruhwirth-schnatterFiniteMixtureMarkov2006}. However, the transition matrix has an exceptionally large dimension ($2^{N}X2^{N}$) {\color{black}which necessitates some changes to the inferential procedure compared to more traditional Markov switching models \citep{fruhwirth-schnatterFiniteMixtureMarkov2006}}, as we will discuss next.

The likelihood of $\bm{v}=(\bm{\theta},\bm{\beta})^T$, where $\bm{\theta}=(\zeta_0,\eta_0,\bm{\zeta},\bm{\eta},{\color{black}\zeta_{0}^{(c)},\eta_{0}^{(c)},\bm{\zeta^{(c)}},\bm{\eta^{(c)}}})^T$ {\color{black}and $\bm{\beta}=(\bm{\delta},\bm{\gamma})$}, given $\bm{y}=\{\bm{y_t}\}_{t=1}^{T}$ and $\bm{S}$ is given by, 
\begin{align}
\text{L}(\bm{y},\bm{S}|\bm{v}) = \prod_{i=1}^{N} \prod_{t=1}^{T} p(y_{it}|S_{it},\bm{y^{(t-1)}},\bm{\beta})\prod_{i=1}^{N} p(S_{i0}) \prod_{t=1}^{T} p(S_{it}|\bm{S_{t-1}},\bm{y^{(t-1)}},\bm{\theta}). \label{eqn:like}
\end{align} When $N$ is large it is not possible to marginalize out $\bm{S}$ from (\ref{eqn:like}), as doing so requires matrix multiplication with $\Gamma(\bm{S_{t}}|\bm{y^{(t-1)}})$ \citep{fruhwirth-schnatterFiniteMixtureMarkov2006}. Therefore, we estimate the unknown elements of $\bm{S}$ along with $\bm{v}$ by sampling both from their joint posterior distribution which, from Bayes' theorem, is proportional to, 
\begin{align}
p(\bm{S},\bm{v}|\bm{y}) \propto \text{L}(\bm{y},\bm{S}|\bm{v})p(\bm{v}), \label{eqn:joint_post}
\end{align} where $p(\bm{v})$ is the prior distribution of $\bm{v}$. We specify  independent uninformative normal and gamma priors for all low-level parameters. As the joint posterior is not available in closed form, we resort to Markov chain Monte Carlo methods, in particular, we used a hybrid Gibbs sampling algorithm with some steps of the Metropolis-Hastings algorithm to sample from it. We sampled most elements of $\bm{v}$ individually, using an adaptive random walk Metropolis step \citep{shabyExploringAdaptiveMetropolis2010}.  The parameter vectors $(\eta_0,{\color{black}\eta_{0}^{(c)},\bm{\eta^{(c)}}})^T$ and $(\zeta_0,{\color{black}\zeta_{0}^{(c)},\bm{\zeta^{(c)}}})^T$ showed high posterior correlations and, therefore, we jointly sampled them using {\color{black}automated factor slice sampling \citep{tibbitsAutomatedFactorSlice2014}}. This doubled the efficiency (minimum effective sample size per hour) of the Gibbs sampler in our application.

It is straightforward to sample each unknown element of $\bm{S}$ one at a time. When $y_{it}=0$, the full conditional of $S_{it}$ is given by,
\begin{align}
\begin{split}
&P\left(S_{it}=1 |\bm{y},\bm{v},\{S_{jt}\}_{j \neq i, t \in \{0,...,T\} }\right) \\
&= \frac{P(y_{it}=0|S_{it}=1)P(S_{it}=1|\bm{S_{t-1}})\prod_{j=1}^{N}p(S_{j(t+1)}|S_{it}=1,\bm{S_{(-i)(t)}})}{\sum_{s_{it}=0}^{1}P(y_{it}=0|S_{it}=s_{it})P(S_{it}=s_{it}|\bm{S_{t-1}})\prod_{j=1}^{N}p(S_{j(t+1)}|S_{it}=s_{it},\bm{S_{(-i)(t)}})}, \label{eqn:one_at_a_time}
\end{split}
\end{align} where $P(y_{it}=0|S_{it}=s_{it})={\color{black}\left(\frac{r_{it}}{r_{it}+\lambda_{it}}\right)^{r_{it}s_{it}}}$ and the dependence of the densities on $\bm{y^{(t-1)}}$, $\bm{\beta}$ and $\bm{\theta}$ are suppressed to reduce the size of the equation. However, one at a time sampling is known to lead to poor mixing in the Markov switching literature due to the strong posterior correlations between the unknown state indicators \citep{scottBayesianMethodsHidden2002}. Better mixing can be achieved by sampling all of $\bm{S}$ jointly from $p(\bm{S}|\bm{y},\bm{v})$ \citep{chibCalculatingPosteriorDistributions1996}. However, this is not computationally feasible with our model when $N$ is large, as it again involves matrix multiplication with $\Gamma(\bm{S_{t}}|\bm{y^{(t-1)}})$ \citep{fruhwirth-schnatterFiniteMixtureMarkov2006}. As an alternative, we propose to block sample $\bm{S}$ with each block containing all the state indicators in a different collection of locations. More specifically, assume we have put the locations into $c=1,...,C$ blocks with $n_c$ locations in block $c$ and that $\sum_{c=1}^{C}n_c=N$. Let $\bm{S_{(c)}}$ be the set of all state indicators in block $c$ and let $\bm{S_{(-c)}}$ be the set of all state indicators outside of block $c$. The idea is to jointly sample $\bm{S_{(c)}}$ from its full conditional distribution, given by, 
\begin{align}
\begin{split}
p(\bm{S_{(c)}}|\bm{S_{(-c)}},\bm{y},\bm{v}) &= p(\bm{S_{(c)T}}|\bm{S_{(-c)(0:T)}},\bm{y},\bm{v}) \\
& \qquad \times \prod_{t=0}^{T-1} p(\bm{S_{(c)t}}|\bm{S_{(c)t+1}},\bm{S_{(-c)(0:t+1)}},\bm{y_{(1:N)(1:t)}},\bm{v}). \label{eqn:bFFBS} 
\end{split}
\end{align} A blocked forward filtering backward sampling (bFFBS) algorithm is needed to sample from (\ref{eqn:bFFBS}), which we provide in the {\color{black} Supplementary Material (SM)} Section 1. The bFFBS algorithm only requires matrix multiplication with a $2^{n_c}X2^{n_c}$ matrix. Our algorithm can be seen as an extension of the individual forward filtering backward sampling (iFFBS) algorithm recently proposed  by \cite{touloupouScalableBayesianInference2020}, who considered $n_c=1$ for all $c$.

Our hybrid Gibbs sampler was implemented using the R package Nimble \citep{valpineProgrammingModelsWriting2017}. Nimble comes with built in Metropolis-Hastings, {\color{black}automated factor slice} and binary (equivalent to one at a time sampling for the unknown presence/absence indicators) samplers. The bFFBS samplers were implemented using Nimble’s custom sampler feature. All Nimble R code, including for the custom bFFBS samplers, are provided {\color{black}on github} (\url{https://github.com/Dirk-Douwes-Schultz/ZS_CMSP_code}). Nimble was chosen as it is extremely fast (C++ compiled) and only requires the coding of new samplers. In the SM Section 2, we provide a simulation study which shows that our proposed Gibbs sampler can recover the true parameters of the {\color{black}ZS-CMSNB} model.

\subsection{Temporal Prediction}

For arbitrary $K$ step ahead temporal predictions, we use a simulation procedure \citep{fruhwirth-schnatterFiniteMixtureMarkov2006} to draw samples from the posterior predictive distributions. Algorithm 1 in the SM will obtain realizations from the posterior predictive distribution of the cases, $y_{i(T+k)}^{[m]} \sim p(y_{i(T+k)}|\bm{y})$, and the presence of the disease, $S_{i(T+k)}^{[m]} \sim p(S_{i(T+k)}|\bm{y})$, for $i=1,...,N$, $k=1,...,K$ and $m=M+1,...,Q$, where the superscript $[m]$ denotes a draw from the posterior distribution of the variable, $M$ is the size of the burn-in sample and $Q$ is the total MCMC sample size. However, as $S_{i(T+k)}^{[m]}$ can only take two values, 0 or 1, it is difficult to interpret the uncertainty around this prediction for the presence of the disease. Therefore, instead of using summaries of $S_{i(T+k)}^{[m]}$ we use summaries of $P(S_{i(T+k)}=1|S_{i(T+k-1)}^{[m]},\bm{S_{(-i)(T+k-1)}^{[m]}},\bm{y_{T+k-1}^{[m]}},\bm{\theta^{[m]}})$. See the SM Section 3 for the Montle Carlo approximations of the posterior predictive distributions.

\subsection{Fitted Values}

Comparing the predictions, for a fixed K, to the observed values is not practical, as it requires fitting the model to multiple time points. This would result in long computational times, with MCMC methods, and unstable estimates for earlier times. Assume in this section $t \in \{1,...,T\}$ and $i \in \{1,...,N\}$. There are three types of fitted values for a Markov switching model: one step ahead, filtered and smoothed \citep{hamiltonEstimationInferenceForecasting1993}. In our Bayesian setting, the smoothed fitted values are given by,
\begin{align}
&S_{it}^{*[m]}=S_{it}^{[m]} \text{ drawn from } p(S_{it}|\bm{y}), \\ 
&y_{it}^{*[m]} \text{ drawn from } p(y_{it}|S_{it}^{[m]},\bm{y^{(t-1)}},\bm{\beta^{[m]}}),
\end{align} for $m=M+1,...,Q$. Note, $y_{it}^{*[m]} \sim p(y_{it}^{*}|\bm{y})$, where $y_{it}^{*}|S_{it},\bm{y^{(t-1)}},\bm{\beta} \sim {\color{black}NB}(S_{it}\lambda_{it},{\color{black}r_{it}})$. That is, $y_{it}^{*}$ represents a new case count generated by the same parameters, state and past counts that generated $y_{it}$. The smoothed fitted value of the state is drawn automatically by the Gibbs sampler used to fit the model, which is an advantage in using data augmentation to fit a Markov switching model. Note that $P(S_{it}=1|\bm{y}) \approx \frac{1}{Q-M} \sum_{m=M+1}^{Q} S_{it}^{[m]}$. When $y_{it}=0$, the $P(S_{it}=1|\bm{y})$ represents the posterior probability the disease is present and thus undetected. Therefore, a map of the posterior mean of $S_{it}$ for $i$ where $y_{it}=0$ can be used by policy makers as a warning system to identify areas reporting 0 cases that have a high chance of the disease being undetected.

Clearly, smoothed fitted values represent fits given all information in the data about the states, including from present and future counts. If the model is not good at predicting the state one step ahead, which is largely based on the estimated transition probabilities of the Markov chain, then the smoothed fitted values may not reveal that as much of the information about $S_{it}$ can come from $y_{it}$. The one step ahead distribution of the states is given by, $p(\bm{S_t}|\bm{y^{(t-1)}},\bm{v})$, and of the cases is given by, $p(\bm{y_t}|\bm{y^{(t-1)}},\bm{v})=\sum_{\bm{s_t} \in \{0,1\}^{N}}p(\bm{y_t}|\bm{S_t}=\bm{s_t},\bm{y^{(t-1)}},\bm{\beta})P(\bm{S_t}=\bm{s_t}|\bm{y^{(t-1)}},\bm{v})$ \citep{fruhwirth-schnatterFiniteMixtureMarkov2006}. As $\bm{v}$ is unknown, we can draw from the posterior of the one step ahead distributions. That is, the one step ahead fitted value of the state is given by,  $S_{it}^{1*[m]}$ drawn from $p(\bm{S_t}|\bm{y^{(t-1)}},\bm{v^{[m]}})$, and of the counts is given by, $y_{it}^{1*[m]}$  drawn from $p(y_{it}|S_{it}^{1*[m]},\bm{y^{(t-1)}},\bm{\beta^{[m]}})$, for $m=M+1,...,Q$, where we use the superscript $1*[m]$ to denote a draw from the posterior of the one step ahead distributions. The advantage of the one step ahead fitted values is that information from $(\bm{y_t},...,\bm{y_{T}})^{T}$ is only used to learn about $\bm{v}$ and not the states. Therefore, they should more accurately reflect the predictions compared to the smoothed values and should be able to diagnosis a lack of fit in the Markov chain component of the model. The distribution $p(\bm{S_t}|\bm{y^{(t-1)}},\bm{v^{[m]}})$ can be calculated using a Hamiltonian forward filter \citep{fruhwirth-schnatterFiniteMixtureMarkov2006} algorithm. However, the filter requires matrix multiplication with $\Gamma(\bm{S_{t}}|\bm{y^{(t-1)}})$ and so is not computationally feasible to run when $N$ is large. 

As an alternative for large $N$, we propose coupled one step ahead fitted values. Assume we wish to calculate the one step ahead fitted values for location $i$ when N is large. We will put $i$ in a location block $c$ with neighboring locations. Then, borrowing notation from (\ref{eqn:bFFBS}), the coupled one step ahead fitted values are given by,
\begin{align}
&S_{it}^{c1*[m]} \text{ drawn from } p(S_{it}|\bm{S_{(-c)(0:t)}^{[m]}},\bm{y^{(t-1)}},\bm{v^{[m]}}), \label{eqn:pres_cone}\\
&y_{it}^{c1*[m]} \text{ drawn from } p(y_{it}|S_{it}^{c1*[m]},\bm{y^{(t-1)}},\bm{\beta^{[m]}}) \label{eqn:case_cone},
\end{align} for $m=M+1,...,Q$, where we use the superscript $c1*[m]$ to denote a draw from the posterior of the coupled one step ahead distributions. The distribution $P(S_{it}=1|\bm{S_{(-c)(0:t)}^{[m]}},\bm{y^{(t-1)}},\bm{v^{[m]}})$ can be calculated using the forward filtering part of the bFFBS algorithm given in the supplementary material. We do not use quantiles of $S_{it}^{c1*[m]}$ as it can only take the two values 0 or 1. Instead we use quantiles of $P(S_{it}=1|\bm{S_{(-c)(0:t)}^{[m]}},\bm{y^{(t-1)}},\bm{v^{[m]}})$, which represents a draw from the posterior distribution of the coupled one step ahead fitted probability of disease presence. We only use information from $(\bm{y_t},...,\bm{y_{T}})^{T}$ to learn about $\bm{S_{(-c)(0:t)}}$. Therefore, if we block neighboring locations together there should be minimal information from $(\bm{y_t},...,\bm{y_{T}})^T$ used in the state estimation.

\section{Analysis of the dengue fever data in Rio de Janeiro}

\subsection{Model Specification and Fitting}

The {\color{black}ZS-CMSNB} model has two components, a nonhomogeneous Markov chain which switches dengue between periods of presence and absence through modeling the reemergence and persistence of the disease, and a {\color{black}negative binomial} component that generates the reported cases when dengue is present. For the {\color{black}negative binomial} component, we use an endemic/epidemic specification for $\lambda_{it}$ \citep{bauerStratifiedSpaceTime2018}, as it can account for temporal autocorrelation due to the onward transmission of the disease, 
\begin{align}
\lambda_{it} &= \lambda_{it}^{AR} y_{i(t-1)}+\lambda_{it}^{EN}. \label{eqn:AR_EN}
\end{align} The auto-regressive rate $\lambda_{it}^{AR}$ is meant to represent the transmission intensity of dengue in district $i$ during month $t$. We model $\lambda_{it}^{AR}$ as, $\log \left(\lambda_{it}^{AR}\right) = \beta_{0i}^{AR}+\beta_{1} \text{Rain}_{t-1} + \beta_2  \text{Temp}_{t-1}$, where $\text{Temp}_{t-1}$ is the maximum temperature in Rio during the previous month, $\text{Rain}_{t-1}$ is the millimeters of rainfall and $\beta_{0i}^{AR} \sim N(\beta_0^{AR},\sigma_{AR})$ are random intercepts accounting for between district differences in transmission intensity, where $\sigma_{AR}$ is the standard deviation of the random effects. Rainfall and temperature are entered in a lagged manner since there is usually a delay separating changes in mosquito population and dengue incidence \citep{coutinhoThresholdConditionsNonAutonomous2006}. The endemic risk $\lambda_{it}^{EN}$ accounts for incidence not due to within area transmission. We compared both space-time and space varying endemic risk using the widely applicable information criterion (WAIC) \citep{gelmanUnderstandingPredictiveInformation2014} (results not shown) and settled on a purely spatial varying risk, $\log \left(\lambda_{it}^{EN}\right) = \beta_{0i}^{EN}$, where $\beta_{0i}^{EN}\sim N(\beta_{0}^{EN},\sigma_{EN})$. {\color{black}Examining the reported dengue counts it appears the overdispersion varies by district, therefore, we specified $r_{it}=r_i$. We used the weakly informative log-normal prior from \cite{bauerStratifiedSpaceTime2018} for $r_i$ which assumes the overdispersion is likely between 1.5 and 5 times the mean count in the district.}

We considered $\bm{z_{it}}=(\text{pop}_{i},\text{HDI}_i,\text{Rain}_{t-1},\text{Temp}_{t-1},\log(y_{i(t-1)}+1))^T$, where $\text{pop}_{i}$ is the population of district $i$ and $\text{HDI}_{i}$ is the human development index of district $i$, as covariates possibly affecting the reemergence and persistence (see equations (\ref{eqn:reemergence}) and (\ref{eqn:persistence})) of dengue. The monthly rainfall and temperature data were obtained from the National Institute for Space Research (\url{http://bancodedados.cptec.inpe.br/}) and the district level human development indexes were obtained from ipeadata (\url{http://www.ipeadata.gov.br/Default.aspx}). Since we condition on the first observation, we specified $p(S_{i1}) \sim Bern(.5)$ if $y_{i1}=0$ and $p(S_{i1})$ is degenerate 1 if $y_{i1}>0$, as the initial state distributions. 

{\color{black}As for covariates potentially associated with the effects of dengue spreading from a neighboring district, i.e. $\bm{z_{01,ijt}^{(c)}}$ and $\bm{z_{11,ijt}^{(c)}}$ in (\ref{eqn:coupled_params}), we considered; $\log(y_{j(t-1)}/pop_{j}+1)$, the prevalence of the disease in the neighboring district; $\log(pop_i\times pop_j)$, a gravity term reflecting the fact that individuals are more likely to move between high population areas; $|NE(j)|$, the number of neighbors of area $j$ which can account for the effects of disease spread potentially being proportioned among neighbors; and, finally, $\log(y_{i(t-1)}+1)$ for the persistence only, to potentially account for crowding out effects. Due to the high computational cost per covariate added to the coupling parameters we started with a simple homogeneous model of disease spread and added coupling covariates sequentially, removing them if they did not improve the WAIC substantially (by 10  units); this process is illustrated in Table \ref{tab:Table 1}. Based on Table \ref{tab:Table 1} we decided on $\bm{z_{01,ijt}^{(c)}}=(\log(y_{j(t-1)}/pop_{j}+1),\log(pop_i\times pop_j))^T$ and a homogeneous effect of neighboring disease spread for the persistence of the disease. {\color{black}That is, we let $\phi_{01,j \to i}^{(t-1) \to t}=\zeta_{0}^{(c)}+\zeta_{1}^{(c)}\log\left(y_{j(t-1)}/pop_j+1\right)+\zeta_{2}^{(c)}\log\left(pop_{i}\times pop_j\right)$ and $\phi_{01,j \to i}^{(t-1) \to t}=\eta_{0}^{(c)}$ for the final model.}}

{\color{black}Each ZS-CMSNB model in Table \ref{tab:Table 1}, and the ZS-CMSP model as it is a special case,} was fit using our proposed Gibbs sampler from Section 3, to the monthly Rio dengue cases for 2011-2017 ($t=1,...,84$) and all districts in the city ($i=1,...,160$). We ran the Gibbs sampler for {\color{black}80,000} iterations on 3 chains with a initial burn in of {\color{black}30,000} iterations. All sampling was started from random values in the parameter space to avoid convergence to local modes. Convergence was checked using the Gelman-Rubin statistic (all estimated parameters $<$1.05) and the minimum effective sample size ($>$1000) \citep{plummerCODAConvergenceDiagnosis2006}. {\color{black}For the final bolded model in Table \ref{tab:Table 1}}, we compared the efficiency (minimum effective sample size per hour) of 3 candidate samplers for the state indicators: a one at time sampler (binary) and two bFFBS samplers with block sizes 1 (iFFBS) and 2 (bFFBS2). For the bFFBS2 sampler, we blocked 71 neighboring locations together leaving 15 locations that could not be matched to a neighbor that were put in single location blocks (remaining 3 locations had all positive observations and did not need to be sampled). The iFFBS sampler was {\color{black}10\%} more efficient than the binary sampler and the bFFBS2 sampler was {\color{black}5\%} more efficient. {\color{black}Therefore, there does not appear to be much if any gain to joint sampling the hidden states in our application.}

{\color{black}We compared the fit of our model to a nested model without the Markov chain, i.e. a model for which $\zeta_0=\eta_0$, $\bm{\zeta}=\bm{\eta}$, $\zeta_0^{(c)}=\eta_0^{(c)}=0$ and $\bm{\zeta^{(c)}}=\bm{\eta^{(c)}}=0$ in equations (\ref{eqn:reemergence})-(\ref{eqn:coupled_params}). Note that this is a standard zero inflated negative binomial (ZINB) model \citep{greeneAccountingExcessZeros1994}. However, the ZINB model does not account for spatio-temporal correlations in the presence of the disease. Therefore, we also compared to a model like the one used by \cite{hoefSpaceTimeZeroinflated2007}. Let $p_{it}=P(S_{it}=1|\bm{\theta_{RE}})$, where $\bm{\theta_{RE}}$ is a vector of random effects, then \cite{hoefSpaceTimeZeroinflated2007} fit a model where $\text{logit}(p_{it})=\alpha_0+\bm{q_{it}^T}\bm{\alpha}+\eta_t+\delta_{it}$, where $\bm{q_{it}}$ is a vector of space time covariates, $\eta_t$ is an overall time trend which followed an autoregressive (AR) process and $\left\{\delta_{it}\right\}_{i=1}^{N}$ followed a proper conditional autoregressive (pCAR) distribution for each time period that was independent across time. As most excess variation in spatio-temporal infectious disease counts occurs temporally and not spatially \citep{bauerBayesianPenalizedSpline2016}, we flip their proposal so that $\text{logit}(p_{it})=\alpha_0+\bm{q_{it}^T}\bm{\alpha}+\eta_i+\delta_{i}(t)$, where $\eta_{i}$ is an overall spatial effect and $\delta_{i}(t)$ represents a time trend for each area. We assume a pCAR distribution for $\left\{\eta_i\right\}_{i=1}^{N}$. \cite{torabiZeroinflatedSpatiotemporalModels2017} suggests using an AR process or splines to model $\delta_{i}(t)$. However, we prefer using autoregression for the spatial-temporal interaction, combining the proposals of \cite{fernandesModellingZeroinflatedSpatiotemporal2009} and \cite{yangMarkovRegressionModels2013}, as it is computationally much simpler and also epidemiologically motivated. That is, we let $\delta_{i}(t)=\rho_1 \log(y_{i(t-1)}+1) +  \rho_2 S_{i(t-1)}$. The motivation being that for the disease to go extinct all previous cases must fail to pass on the disease but since $y_{i(t-1)}$ represents reported cases there might still be an effect at $y_{i(t-1)}=0$ due to the disease being undetected. We take $\bm{q_{it}}=\{\bm{z_{it}}\} \setminus \{\log(y_{i(t-1)}+1)\}$ as $\log(y_{i(t-1)}+1)$ is already included in the model. We call this model a zero inflated negative binomial random effect (ZINBRE) model.

Finally, we also compared the fit of our model to a model without zero inflation (endemic/epidemic) and to a Poisson version of our model (ZS-CMSP). Table \ref{tab:Table 1} gives the results of the model comparison using WAIC for the five classes of models. {\color{black}The endemic/epidemic, ZINB and ZINBRE models in Table \ref{tab:Table 1} were fit using standard MCMC methods in Nimble, the code is available on github (\url{https://github.com/Dirk-Douwes-Schultz/ZS_CMSP_code}).} From Table \ref{tab:Table 1}, incorporation of zero inflation and overdispersion significantly improves model fit, and the ZS-CMSNB model is the superior model for the zero inflation.}

\renewcommand{\arraystretch}{1.5}

\begin{table}[t]
\centering
\caption{\label{tab:Table 1}{\color{black}Model comparison using WAIC for models fitted to the dengue data. {\color{black}Note, $\emptyset$ refers to no covariates, or a homogeneous model of disease spread, that its, $\phi_{01,j \to i}^{(t-1) \to t}=\zeta_{0}^{(c)}$ or $\phi_{11,j \to i}^{(t-1) \to t}=\eta_{0}^{(c)}$.} {\color{black}The bolded model is our final ZS-CMSNB model as we do not accept models that are more complex and do not improve the WAIC substantially (by 10 units).} NP = Neighboring prevalence $=\log(y_{j(t-1)}/pop_{j}+1)$, G = Gravity term $=\log(pop_i\times pop_j)$, LC = Local cases $=\log(y_{i(t-1)}+1)$, NN = Number of neighbors $=|NE(j)|$.}}
{\color{black}
\begin{tabular}{llll}
\hline
\bf{Model} & $\bm{z_{01,ijt}^{(c)}}$  & $\bm{z_{11,ijt}^{(c)}}$ & \bf{WAIC}  \\ 
\hline
endemic/epidemic & -- & -- & 69,939    \\ 
ZINB & -- & -- &67,798     \\
ZINBRE & -- & -- & 67,782   \\
 ZS-CMSNB & $\emptyset$  & $\emptyset$ & 67,729  \\
  &$\left(\text{NP}\right)^T$  & $\emptyset$ & 67,652  \\
  & {\boldmath$\left(\text{\bf{NP}}, \text{\bf{G}}\right)^T$}  & $\bm{\emptyset}$ & \bf{67,632}  \\
  & $\left(\text{NP},\text{G}\right)^T$ & $\left(\text{NP}\right)^T$ & 67,631 \\
  & $\left(\text{NP},\text{G}\right)^T$ & $\left(\text{NP},\text{LC}\right)^T$  & 67,628   \\
   & $\left(\text{NP},\text{G},\text{NN}\right)^T$  & $\emptyset$ & 67,648  \\
   ZS-CMSP & $\emptyset$  & $\emptyset$ & 90,458 \\ \hline
\end{tabular}
}
\end{table} 

{\color{black}\subsection{Results}}

{\color{black} From Table \ref{tab:Table 1}, the most relevant factors for determining the strength of dengue spread from a neighboring area are, the reported prevalence of dengue in the neighboring area, and the population of the neighboring area and the home area.  {\color{black}To quantify the association between the effects of neighboring dengue spread and these two factors we can use the odds ratio (OR) of dengue reemergence given disease spread from a neighboring area, or mathematically, {\small \begin{align} \frac{\Omega(S_{it}=1|S_{i(t-1)}=0,S_{j(t-1)}=1)}{\Omega(S_{it}=1|S_{i(t-1)}=0,S_{j(t-1)}=0)}=\exp\left(\zeta_{0}^{(c)}+\zeta_{1}^{(c)}\log(y_{j(t-1)}/pop_j+1)+\zeta_{2}^{(c)}\log(pop_{i}\times pop_j)\right),\label{eqn:OR_spread}\end{align}}for $j \in NE(i)$,  where $\Omega(A)=P(A)/(1-P(A))$. From Table \ref{tab:Table 3}, which shows estimates from the Markov chain component of the model, the posterior mean and 95\% posterior credible interval of $\zeta_{1}^{(c)}$ is 5.63 {[}3.33-8.67{]} and of $\zeta_{2}^{(c)}$ is .22 {[}.09-.36{]}. Therefore, the effect of dengue spreading from a neighboring area is positively associated with the reported prevalence of dengue in the neighboring area and the population of both areas, especially with the prevalence. To help visualize these associations we have plotted, in Figure \ref{fig:OR_spread}, (\ref{eqn:OR_spread}) versus the population of the neighboring area (top) and the reported cases of dengue in the neighboring area (bottom).}
\begin{figure}[t]
 	\centering
 	\includegraphics[width=.75\textwidth]{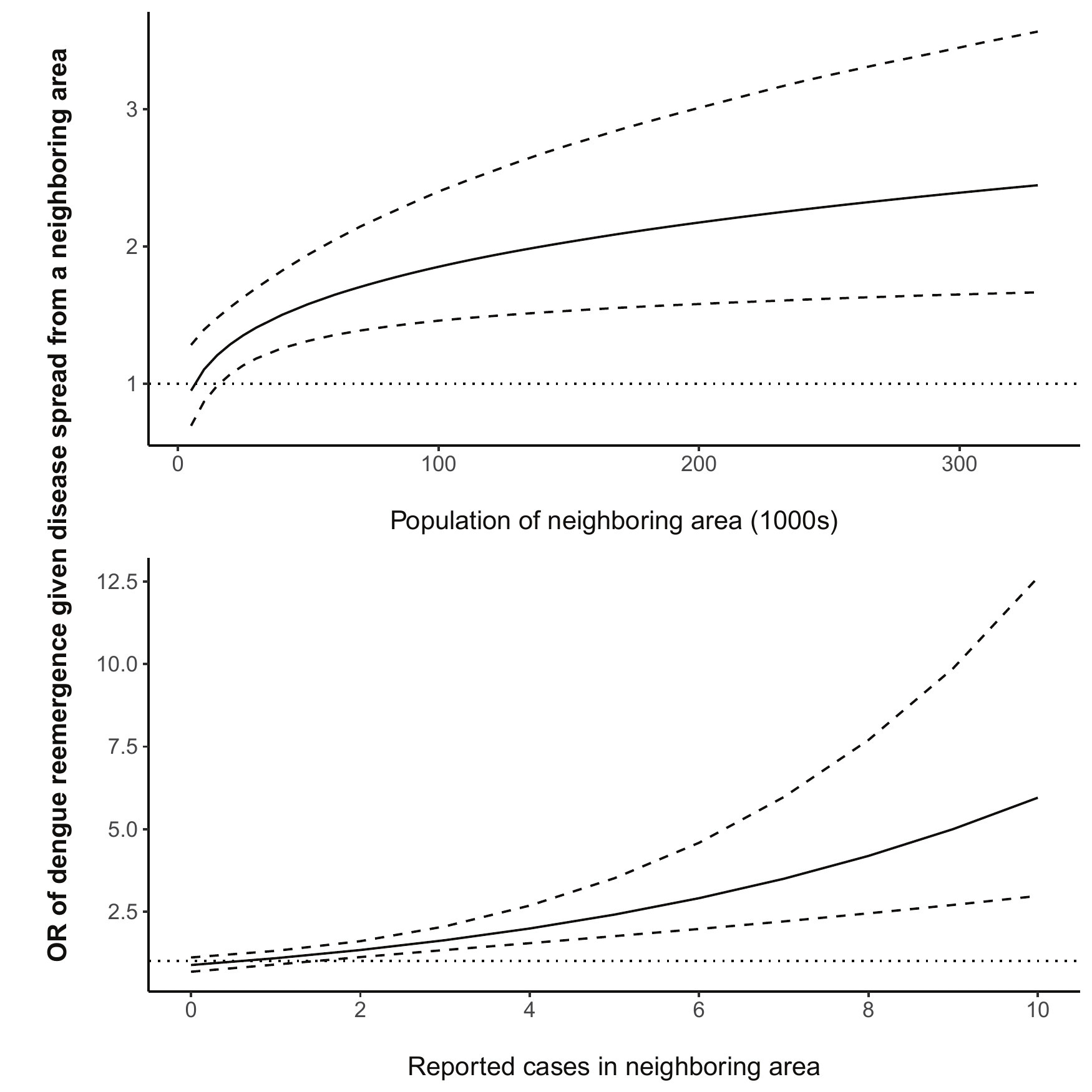}
	\caption{ \label{fig:OR_spread} {\color{black}Posterior means (solid lines) and 95\% posterior credible intervals (dashed lines) of (top) $\exp\left(\zeta_{0}^{(c)}+\zeta_{1}^{(c)}\log(.08+1)+\zeta_{2}^{(c)}\log(10\times pop_j)\right)$ versus $pop_j$ and (bottom) $\exp\left(\zeta_{0}^{(c)}+\zeta_{1}^{(c)}\log(cases_j/26+1)+\zeta_{2}^{(c)}\log(10\times 26)\right)$ versus $cases_j$ {\color{black} (see (\ref{eqn:OR_spread})), where, 10 is the population of a small area, 26 is the median population and .08 (per 1000) is the median prevalence of dengue.}}}
\end{figure}  In the plots, we fixed the population of the home area at 10,000, to represent a small area, and for the top graph we fixed prevalence in the neighboring area at its median value (.08 per 1000), and for the bottom graph we fixed the population in the neighboring area at its median value (26,000) (exact formulas are given in the caption). From Figure \ref{fig:OR_spread} (top), the effect of dengue spreading from a neighboring area increases gradually with the population of the neighboring area and there is likely not a large effect of disease spread from low population areas, at least at median prevalence. These estimates are reasonable as there will be much more travel to high population neighbors compared to low population neighbors. From Figure \ref{fig:OR_spread} (bottom), the effect of dengue spreading from a neighboring area increases rapidly with the reported prevalence of the disease in the neighboring area. At a prevalence of 10/26 = .38, where we have cut off the graph to improve visualization, the odds ratio is 5.95 [2.97-12.62], meaning the odds of dengue reemergence is increased 6 (likely lowest 3) times due to the disease spread, a large effect. A prevalence of .38 is about at epidemic levels, 75th percentile, and, therefore, the figure implies that if an epidemic occurs in an area there is a good chance that dengue will spread to neighboring areas where it is absent.

We also wanted to investigate whether dengue spreads between areas undetected. Note that the OR of dengue reemergence given disease spread from a neighboring area where dengue is undetected is given by $\exp\left(\zeta_{0}^{(c)}+\zeta_{2}^{(c)}\log(pop_{i}\times pop_j)\right)$. Through repeated calculations we found that, {\color{black} {\em a posteriori}}, this quantity  has a 75\% chance of being greater than one if $pop_j>685.4/pop_i$. Setting $pop_i=10$ (1000) gives $pop_j>68.53$, which represents the top 15 percent of districts in terms of population size. Therefore, the disease likely spreads undetected only from the larger districts.}

\begin{figure}[t]
 	\centering
 	\includegraphics[width=.8\textwidth]{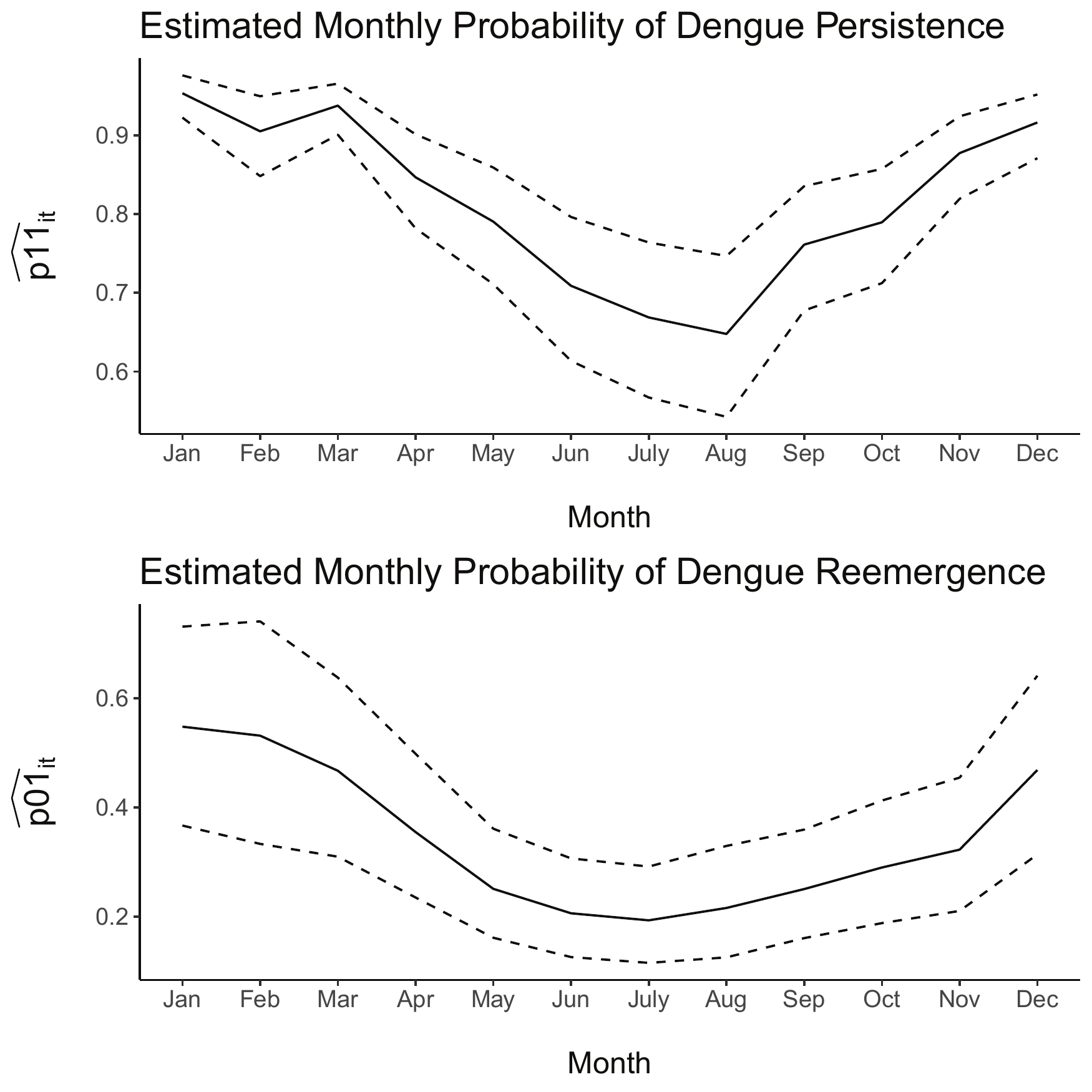}
	\caption{ \label{fig:reem_pers} Posterior means (solid lines) and 95\% posterior credible intervals (dashed lines) of the estimated monthly probabilities of dengue reemergence and persistence in a small district (pop=10,000). We assumed average monthly temperature and rainfall values, and 2 cases {\color{black}(median)} reported in the previous month for the persistence probabilities. } 
\end{figure}

The estimated coefficients for the Markov chain part of the fitted {\color{black}ZS-CMSNB} model are given in Table \ref{tab:Table 3}. From the estimated intercepts, {\color{black}in small areas} consecutive periods of disease presence are heavily favored on average, while consecutive periods of disease absence are only {\color{black}moderately} favored. This highlights the importance of preventing dengue reemergence {\color{black}in small areas} {\color{black}(one effective way would be to prevent disease spread from neighboring epidemic areas as shown in Figure \ref{fig:OR_spread})}, as when the disease does reemerge it will likely persist for some time. {\color{black}We shift the intercepts to small districts (pop=10,000) as that is where most of the 0s occur (there are not many 0s in large districts).} Population size has a significantly larger effect on dengue reemergence compared to persistence {\color{black}(diff=.04 {[}0-.08{]})}{\color{black}, especially considering population increases the effects of dengue spreading from neighboring areas on the reemergence of the disease, see Figure \ref{fig:OR_spread}}. This can explain why, in Figure \ref{fig:avg_time}, there is more spatial variation observed in the average probabilities of dengue reemergence compared to persistence. Additionally, the strong positive association between population size and dengue reemergence means we would expect much longer periods of disease absence in the smaller districts compared to larger districts, which follows the well-known theories from \cite{bartlettMeaslesPeriodicityCommunity1957}. Interestingly, we found no association between human development index and the risk of dengue reemergence or persistence. Both rainfall and temperature have a strong positive association with dengue reemergence and persistence. For example, we estimated that a one degree rise in maximum temperature during the previous month is associated with a {\color{black}24 [11-39]} percent increase in the odds of dengue reemerging and a {\color{black}21 [13-31]} percent increase in the odds of dengue persisting. Therefore, we will have longer periods of disease absence during the winter (at least in the smaller districts) and longer periods of disease presence during the summer, which follows patterns in the mosquito population. {\color{black}Rainfall has a significantly larger effect on the persistence of the disease (diff=.08 {[}.02-.15{]}), which clearly manifests in Figure \ref{fig:reem_pers}, and could be due to the role played by rainfall in the lifecycle of the mosquito egg \citep{coutinhoThresholdConditionsNonAutonomous2006}. When the disease is present water is needed for egg laying and so rainfall helps the mosquito population persist, however, if the disease is absent there are not many mosquitoes and rainfall could wash away vertically infected mosquito eggs.} The effect of the previous months cases, $\log y_{i(t-1)}$, on the persistence of the disease is quite large and reflects the fact that dengue will only go extinct when there is a small number of infected individuals.

\begin{table}[t]
\centering
\caption{\label{tab:Table 3} Posterior means and 95\% posterior credible intervals (in squared brackets) for the estimated parameters from the Markov chain part of the fitted {\color{black}ZS-CMSNB} model. {\color{black}Intercepts shifted so they represent the probabilities of dengue reemergence and persistence with 0 neighbors infected, pop=10,000 and all other covariates fixed at mean or median (if very skewed) values. Other rows represent odds ratios, i.e. $\exp \left(\zeta_k \right)$ and $\exp \left(\eta_k \right)$ for $k=$ pop=1,..., $=\log(y_{i(t-1)}+1)=K$. The reemergence column of the Neighborhood Presence row shows $\exp(\phi_{01,j \to i}^{(t-1) \to t})$ calculated at median values of neighboring pop/prevalence and pop=10,000 for the home area, the persistence column shows $\exp(\eta_0^{(c)})$. {\color{black}The rows beneath the Neighborhood Presence row show, respectively, unmodified estimates of $\zeta_0^{(c)}/\eta_0^{(c)}$, $\zeta_1^{(c)}$ and $\zeta_2^{(c)}$.}}}

\begin{tabular}{lrr} 
 \hline                                    & \multicolumn{2}{c}{\textbf{Probability or Odds Ratio}}            \\ \hline 
                             & \textbf{Reemergence} & \textbf{Persistence}  \\[-5pt] 
\textbf{Covariate}                             &  \textbf{(absence to presence)} &  \textbf{(presence to presence)} \\  \hline
Intercept {\color{black}(shifted)}                            & {\color{black}.33 {[}.22-.46{]}}               & {\color{black}.84 {[}.78-.90{]}}                \\ 
pop (1000s)                           & {\color{black}1.06 {[}1.02-1.11{]}}           & {\color{black}1.02 {[}1.01-1.03{]}}           \\ 
HDI {\color{black}(.1)}                                 & {\color{black}1.07 {[}.71-1.52{]}}            & {\color{black}.82 {[}.62-1.09{]}}             \\ 
$\text{Temp}_{t-1}$ (Celc.)           & {\color{black}1.24 {[}1.11-1.39{]}}           & {\color{black}1.21 {[}1.13-1.31{]}}           \\ 
$\text{Rain}_{t-1}$ (10 $\text{mm}$) & {\color{black}1.04 {[}1-1.09{]}}          & {\color{black}1.13 {[}1.09-1.17{]}}            \\
$\log(y_{i(t-1)}+1)$                   & --                                & {\color{black}5.07 {[}4.17-6.19{]}}            \\
Neighborhood Presence                   & {\color{black}1.35 {[}1.14-1.64{]}}           & {\color{black}1.12 {[}1.01-1.23{]}}          \\[-9pt]  
                &  {\color{black}(average effect in small area)}  &          \\
                {\color{black}$\quad \text{Intercept}$} & {\color{black}-1.38 {[}-2.28- -.51{]}} & {\color{black}.12 {[}.01-.23{]}} \\ 
                {\color{black}$\quad \log\left(y_{j(t-1)}/pop_j+1\right)$} & {\color{black}5.63 {[}3.33-8.67{]}} & -- \\
                {\color{black}$\quad \log\left(pop_i \times pop_j\right)$} & {\color{black}.22 {[}.09-.36{]}} & -- \\\hline
\end{tabular}
\end{table}

From the Neighborhood Presence row of Table \ref{tab:Table 3}, we estimated that dengue being present in a neighboring district during the previous month is associated with an {\color{black}35 [14-64]} percent increase in the odds of dengue reemerging {\color{black}in a small area on average} and a {\color{black}12 [1-23]} percent increase in the odds of dengue persisting. The {\color{black}average} effect of neighboring dengue spread on the reemergence of the disease is much higher than the {\color{black}homogeneous} effect on the persistence. Therefore, efforts to restrict cross infection between districts will {\color{black}typically} be more effective if one district is not already infected. {\color{black}A possible explanation is a "spark effect", like forest fires, in that only a small amount of disease spread could be needed to start the disease being present in an area but will not contribute much to, say, a large epidemic. Note, however, that the Neighborhood presence row of Table \ref{tab:Table 3} only shows the average effect of neighboring dengue spread on the reemergence of the disease in a small area, i.e. $\exp\left(\zeta_{0}^{(c)}+\zeta_{1}^{(c)}\log(.08+1)+\zeta_{2}^{(c)}\log(10\times 26)\right)$, and the actually effect varies quite a lot with space and time as shown in Figure \ref{fig:OR_spread}, making broad comparisons somewhat challenging.}

Figure \ref{fig:reem_pers} shows posterior summaries of the estimated monthly probabilities of dengue reemergence and persistence for a small district (pop=10,000) {\color{black}without neighboring disease spread}. {\color{black}The temporal evolution of the two probabilities are clearly different which mainly reflects the increased effect of rainfall on the persistence of the disease. Additionally, the figure illustrates how the model can recreate the seasonal switching of dengue between long periods of disease absence in the winter and long periods of disease presence in the summer \citep{adamsHowImportantVertical2010}.}

\begin{table}[t]
\centering
\caption{\label{tab:Table 4}Posterior means and 95\% posterior credible intervals (in squared brackets) from the {\color{black}negative binomial} part of the fitted {\color{black}ZS-CMSNB} model. {\color{black}$\overbar{r}=1/N\sum_{i=1}^{N}r_i$ and $\sigma_r=\sqrt{1/(N-1)\sum_{i=1}^{N}(r_i-\overbar{r})^2}$}}
\begin{tabular}{lcr}
 \hline \textbf{Covariate} & \textbf{Parameter} & \textbf{Estimate}   \\ \hline
Intercept AR           & $\beta_0^{AR}$        & {\color{black}-.37 [-.4- -.33]} \\
$\text{Rain}_{t-1}$ (10 mm) AR & $\beta_1$       & {\color{black}.036 [.03-.04]}  \\
$\text{Temp}_{t-1}$ (C) AR & $\beta_2$       & {\color{black}.23 [.22-.24]}  \\
Std. dev AR & $\sigma_{AR}$       &  {\color{black}.12 [.09-.14]}  \\
Intercept EN & $\beta_0^{AR}$      & {\color{black}.37 [.29-.45]}  \\
Std. dev EN & $\sigma_{EN}$       & {\color{black}.42 [.35-.5]}  \\
{\color{black}Avg. overdispersion} & {\color{black}$\overbar{r}$}       & {\color{black}2.27 [2.16-2.39]}  \\
{\color{black}Std. dev overdispersion} & {\color{black}$\sigma_r$}       & {\color{black}1.08 [.92-1.37]}  \\
\hline
\end{tabular}
\end{table} 

Table \ref{tab:Table 4} gives the estimated parameters from the {\color{black}negative binomial} part of the model. When dengue is present in a district, increases in temperature and rainfall lead to increased transmission of the disease. There is significant between district differences in the transmission rate and endemic risk. The differences in the endemic risk could be driven by either differences in the environment or reporting rates. {\color{black}There is significant overdispersion on average (variance is 11 times the mean on average) and there is significant between district differences in the overdispersion.}

\subsection{Fitted Values and Predictions}

\begin{figure}[ht]
 	\centering
 	\includegraphics[width=\textwidth]{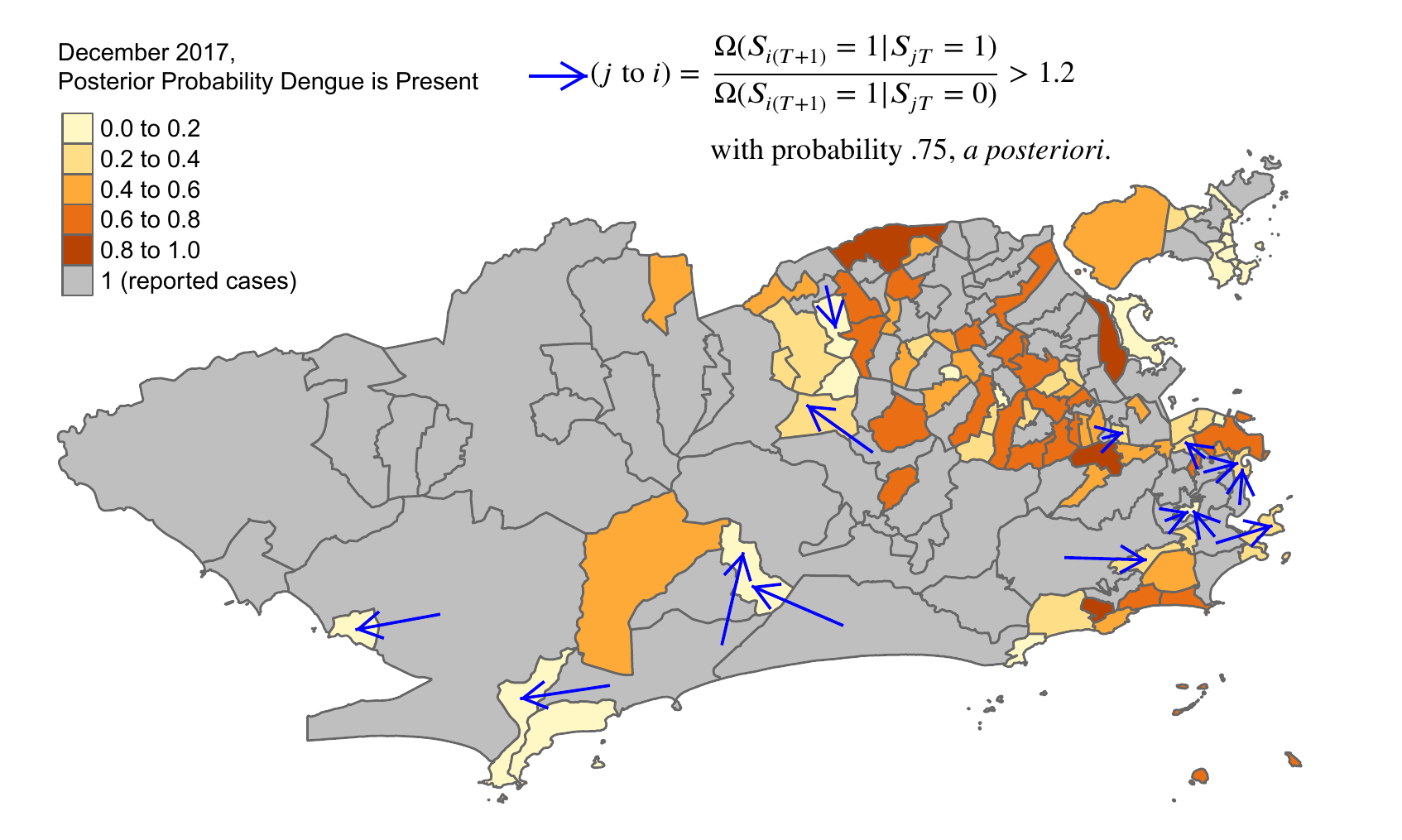}
	\caption{ \label{warn_sys1}  {\color{black}Posterior probability that dengue is present for December 2017 (time $T$). Districts that have reported cases are in grey to distinguish them from districts where dengue may be undetected. A blue arrow is drawn from $j \in NE(i)$ to $i$ if $\frac{\Omega(S_{i(T+1)}=1|S_{jT=1})}{\Omega(S_{i(T+1)}=1|S_{jT=0})}$ is greater than 1.2 with probability .75, \emph{a posteriori}, where $\Omega(A)=P(A)/(1-P(A))$.}} 
\end{figure} 

{\color{black}Figure \ref{warn_sys1} shows a map of the posterior probability that dengue is present in the districts of Rio de Janeiro during December 2017 (time $T$). If $y_{iT}>0$ then the posterior probability that dengue is present in district $i$ during time $T$ is 1 (grey districts on the map). Otherwise, if $y_{iT}=0$ the disease may be undetected and we can approximate $P(S_{iT}=1|\bm{y}) \approx \frac{1}{Q-M}\sum_{m=M+1}^{Q}S_{iT}^{[m]}$. A blue arrow is drawn from $j \in NE(i)$ to $i$ if $\frac{\Omega(S_{i(T+1)}=1|S_{jT=1})}{\Omega(S_{i(T+1)}=1|S_{jT=0})}=S_{iT}\exp\left(\phi_{11,j \to i}^{T \to (T+1)}\right)+(1-S_{iT})\exp\left(\phi_{01,j \to i}^{T \to (T+1)}\right)$ is greater than 1.2 with probability .75, \emph{a posteriori}, where $\Omega(A)=P(A)/(1-P(A))$. That is, if the current effect of disease spread from the neighboring area is likely greater than the effect of a one degree rise in temperature (see Table \ref{tab:Table 3}). The map shows where dengue currently is in the city and where it is spreading to and, therefore, would be of interest to policy makers. For example, the Santa Teresa district has 3 arrows coming out of it indicating it is an important source of dengue spread in the city currently.}

\begin{figure}[ht]
 	\centering
 	\includegraphics[width=16 cm,height=12 cm]{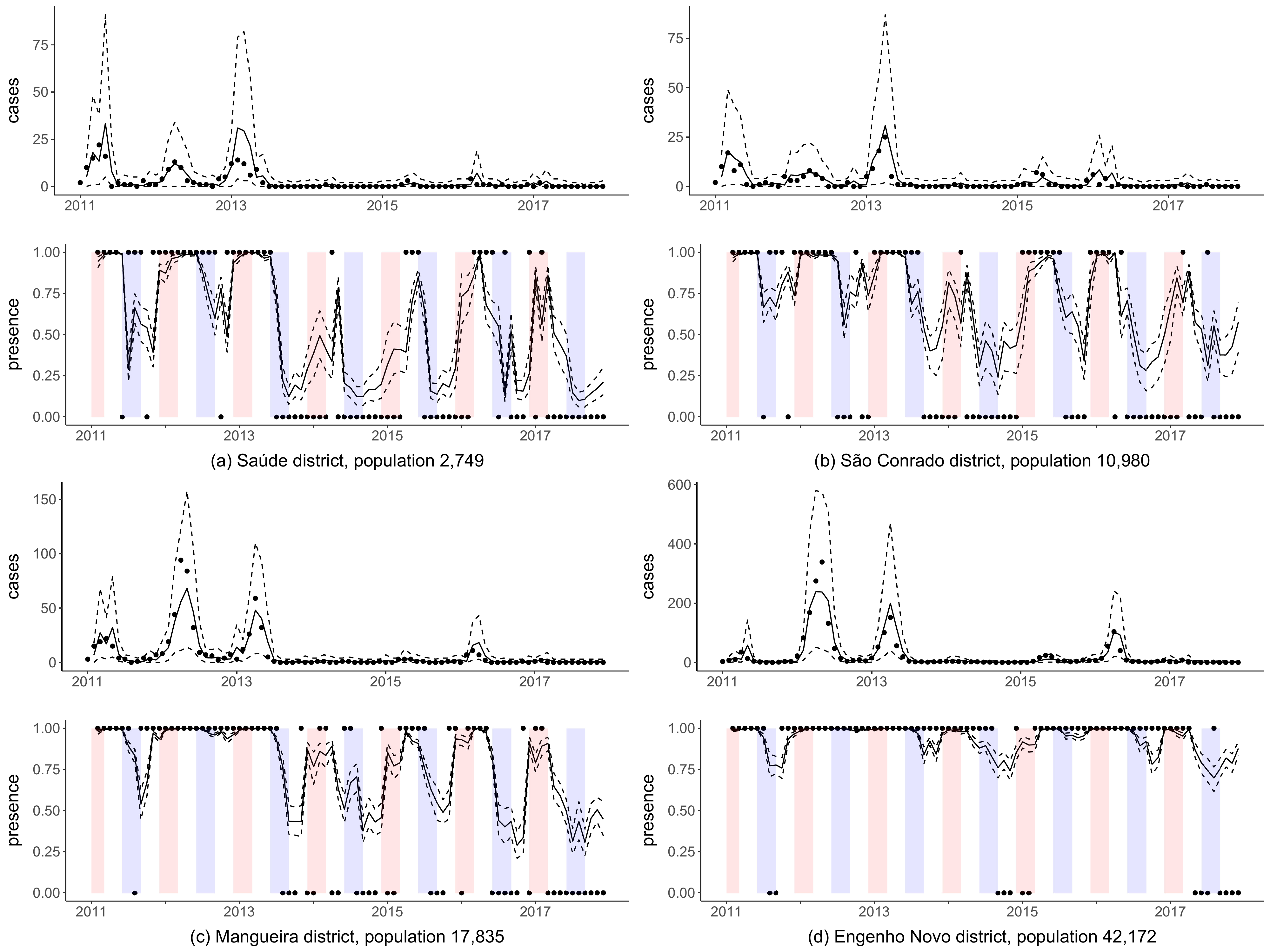}
	\caption{ Posterior summaries of the coupled one month ahead fitted values for 4 districts. (top graphs) Coupled one month ahead fitted values of cases versus observed cases. (bottom graphs) Coupled one month ahead fitted values of dengue presence risk versus observed presence (0=0 reported cases which may not correspond to the actual absence of the disease). Posterior means (solid lines), 95\% posterior credible intervals (dashed lines) and observed (points). Summer/winter seasons highlighted in red/blue in the bottom graphs. \label{fig:fitted}}
\end{figure} 

\begin{figure}[ht]
 	\centering
 	\includegraphics[width=.75\textwidth]{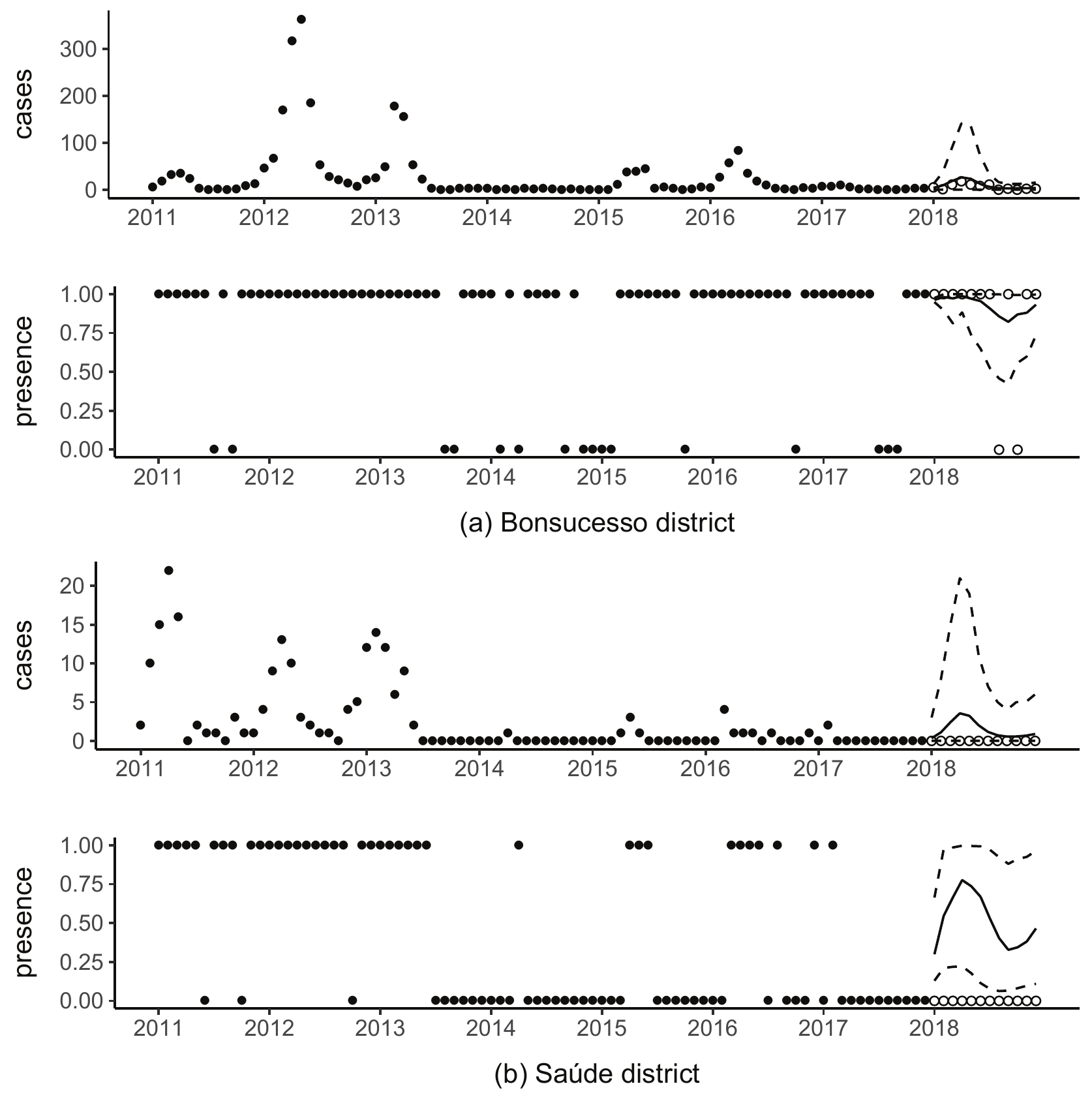}
	\caption{ Summaries of the 1-12 month ahead posterior predictive distributions of the cases (top graphs) and dengue presence risk (bottom graphs) for 2 districts. Posterior predictive means (solid lines) and 95\% posterior predictive credible intervals (dashed lines). Solid circles are from the last {\color{black}7 years (2011-2017)} used to fit the model and open circles are future observed values. \label{fig:pred}} 
\end{figure} 

{\color{black} Panels of Figure \ref{fig:fitted} show} posterior summaries of the coupled one month ahead fitted values, see equations  (\ref{eqn:pres_cone})-(\ref{eqn:case_cone}), of dengue cases (top graphs) and dengue presence (bottom graphs), for 4 districts. These were constructed by running the blocked forward filter in blocks with 2 neighboring locations, with each MCMC draw, as explained in Section 3.2. {\color{black}For Figure \ref{fig:fitted}, we chose mostly small districts that had a good mix of zeros and positive counts to illustrate both the zero inflated and count components of the model.} Generally, the model is able to predict, in sample, the presence of the disease one month ahead well when the disease is observed to be present. It is difficult to assess the presence fits when the disease is reportedly absent, as dengue could be undetected and actually present. {\color{black} The fit to the cases is also reasonable, however, there does appear to be some overestimation of uncertainty at high counts. This is due to the quadratic mean variance relationship of the negative binomial distribution. There is an advantage to this mean variance relationship, however, for estimating the Markov chain part of the model as most of the weight is placed at low counts which is where the reemergence and persistence of the disease occurs. As an additional goodness of fit measure we plot the autocorrelation function of the Pearson residuals in the SM Section 4 for the 4 districts of Figure \ref{fig:fitted}. There are not many significant autocorrelations, {\color{black} suggesting that there is no structure left in the residuals.}} 

{\color{black} Panels of Figure \ref{fig:pred} show} summaries of the 1-12 month ahead posterior predictive distributions of dengue cases (top graphs) and dengue presence risk (bottom graphs) in two districts. These were calculated using Algorithm 1 with $K=12$. {\color{black}There was very little dengue activity in 2018 and we chose to show Bonsucesso as it was one of the few districts that experienced a small epidemic while Sa\'ude represents a more typical district.} We assumed average monthly temperature and rainfall values in the calculations, although other forecasting scenarios could be considered, such as high temperature scenarios, or forecasts of the meteorological variables could be used. Both predictions of dengue presence and the case counts show a clear seasonal pattern, peaking in the summer and declining during the winter. The uncertainty in the predictions of dengue presence is large {\color{black} in the Sa\'ude district}. {\color{black}A reasonable explanation for this large uncertainty is that Sa\'ude reported 0 cases for the last time point, meaning there is uncertainty around whether dengue is present in Sa\'ude at time $T=84$. As Sa\'ude is a small district (pop=2,000) there is a large difference in the probabilities of disease reemergence and persistence (due to the larger effect of population size on the reemergence of the disease), and so uncertainty around the presence of the disease previously leads to wide confidence intervals for the future predictions of disease presence.}

\section{Concluding Remarks}

We have proposed a zero-state coupled Markov switching {\color{black}negative binomial} {\color{black}(ZS-CMSNB)} model for general spatio-temporal infectious disease counts that contain an excess of zeros. {\color{black}Our approach can model nonhomogeneous switching between long periods of disease presence and absence while accounting for space-time heterogeneity in the effects of disease spread between areas.} The main difference with existing {\color{black}ZIC} models used in spatio-temporal disease mapping is that the disease switches between periods of presence and absence in each area through a series of nonhomogeneous Markov chains coupled between neighboring locations, as opposed to the presence of the disease forming a series of conditionally independent Bernoulli random variables, which is presently the most popular approach {\color{black}\citep{youngZeroinflatedModelingPart2020}}. The difference is essentially between a finite mixture and finite Markov mixture model \citep{fruhwirth-schnatterFiniteMixtureMarkov2006}, although we also introduce some spatial dependence in the states by coupling the partially hidden Markov chains between neighboring areas.  

{\color{black}There are several alternatives to our coupled Markov chains, within the framework of finite mixture ZIC models \citep{youngZeroinflatedModelingPart2020}, for accounting for spatio-temporal correlations in the presence of the disease, mainly: random effects \citep{hoefSpaceTimeZeroinflated2007}, splines \citep{ghosalHierarchicalMixedEffect2020} and autoregression \citep{fernandesModellingZeroinflatedSpatiotemporal2009,yangMarkovRegressionModels2013}. The main advantages of our approach are (1) we allow for each covariate, and between area disease spread, to have a separate effect on the reemergence of the disease compared to the persistence, which is epidemiologically justified in many instances and (2) it is much easier to quantify space-time heterogeneity in the effects of disease spread between areas with our approach as we allow the effects of neighboring disease spread to depend on a vector of space-time covariates related to either area. In our application of dengue fever our model allowed for several interesting insights into the epidemiology of the disease beyond what existing ZIC methods can provide, some examples being: dengue is more likely to spread from high population areas controlling for prevalence, if an epidemic of dengue occurs in an area the disease is very likely to spread to neighboring areas, there is typically a larger effect of neighboring dengue spread on the reemergence of the disease compared to the persistence, and rainfall has a larger effect on the persistence of dengue which leads to clear differences in the temporal evolutions of the two probabilities (see Section 4 for these and more examples). Additionally, our model fit the dengue data better than a reasonably specified alternative existing ZIC model that combined random effects and autoregression, although there are many possible specifications for such a model.}

Although we have applied the {\color{black}ZS-CMSNB} model to spatio-temporal infectious disease counts, it could be applied to data in other fields as well. For example, \cite{malyshkinaZerostateMarkovSwitching2010} considered a zero state Markov switching count model, without coupling, for modeling traffic accidents across 335 highway segments in Indiana between 1995-1999. In their application, the zero state represented a low risk of accidents and the count state represented a high risk of accidents. It could be that a highway segment in the high risk state signals that nearby highway segments are also unsafe, and by coupling the chains between neighboring highways we could borrow strength between them to help determine the states.

{\color{black}There are also some limitations with our approach. We use a negative binomial distribution to account for overdispersion, but other count distributions could be used, such as the generalized Poisson distribution \citep{joeGeneralizedPoissonDistribution2005}. The negative binomial distribution places a lot of weight on low counts which is helpful when modeling the reemergence and persistence of a disease, as these events mainly occur at low counts, but has the disadvantage of leading to an overestimation of uncertainty during epidemic periods. Additionally, our approach relies heavily on having good covariates that reflect knowledge of the disease. We could incorporate random effects or splines into $\text{logit}(p01_{it})$ or $\text{logit}(p11_{it})$ to account for additionally heterogeneity when known covariates are not sufficient. Finally,} we are only {\color{black}explicitly} considering disease spread between neighboring areas and, especially in a city, people often move around far outside their neighboring areas. {\color{black}We could allow for disease spread from all areas and incorporate distance into the coupling parameters, although this would likely come at a great computational cost that would need to be overcome to make it a viable approach.} 

\section*{Acknowledgements}
This work is part of the PhD thesis of D. Douwes-Schultz under the supervision of A. M. Schmidt in the Graduate Program of Biostatistics at McGill University, Canada. Schmidt is grateful for financial support from the Natural Sciences and Engineering Research Council (NSERC) of Canada (Discovery Grant RGPIN-2017-04999). {\color{black}Douwes-Schultz is grateful for financial support from IVADO and the Canada First Research Excellence Fund / Apogée (PhD Excellence Scholarship). This research was enabled in part by support provided by Calcul Québec (www.calculquebec.ca) and Compute Canada (www.computecanada.ca).}

\bibliography{ms}

\end{document}


\maketitle

\tableofcontents

\section{The blocked forward filtering backward sampling (bFFBS) algorithm}

Borrowing the notation from the main text, sampling $
\bm{S}$ in the Gibbs sampler involves the following steps. First we initialize $\bm{S}^{[1]}$ by setting $S_{it}^{[1]}=0$ whenever $y_{it}=0$, and we set $S_{it}^{[m]}=1$ whenever $y_{it}>0$ for all $m$, since only the {\color{black}count} process can produce a positive count. Then the following steps are repeated for $m=2,...,Q$, {\color{black} where $Q$ is the total number of iterations},
\begin{enumerate}
    \item Sample $\bm{v}^{[m]}$ from $p(\bm{v}|\bm{S^{[m-1]}},\bm{y})$
    \item Sample $\bm{S_{(c)}^{[m]}}$ from $p(\bm{S_{(c)}}|\bm{S_{(1)}^{[m]}},...,\bm{S_{(c-1)}^{[m]}},\bm{S_{(c+1)}^{[m-1]}},...,\bm{S_{(C)}^{[m-1]}},\bm{v^{[m]}},\bm{y})$ for $c=1,...,C$.
\end{enumerate} Step 1 is broken up into (mostly) Metropolis-Hastings steps as described in the main text. In step 2, we only sample unknown (i.e. when $y_{it}=0$) state indicators in the block. Here we provide the algorithms for sampling from $p(\bm{S_{(c)}}|\bm{S_{(-c)}},\bm{v},\bm{y})$ needed for step 2. \cite{touloupouScalableBayesianInference2020} derived the algorithm for $n_c=1$ for all $c$, and it is straightforward to generalize it to arbitrary $n_c$.

Note that, 
\begin{align}
\begin{split}
p(\bm{S_{(c)}}|\bm{S_{(-c)}},\bm{y},\bm{v}) &= p(\bm{S_{(c)T}}|\bm{S_{(-c)(0:T)}},\bm{y},\bm{v}) \\
& \qquad \times \prod_{t=0}^{T-1} p(\bm{S_{(c)t}}|\bm{S_{(c)t+1}},\bm{S_{(-c)(0:t+1)}},\bm{y_{(1:N)(1:t)}},\bm{v}). \label{eqn:bFFBS} 
\end{split}
\end{align}
and that, 
\begin{align*}
p(\bm{S_{(c)t}}|\bm{S_{(c)t+1}},\bm{S_{(-c)(0:t+1)}},\bm{y_{(1:N)(1:t)}},\bm{v}) &\propto p(\bm{S_{(c)(t+1)}}|\bm{S_{(c)(t)}},\bm{S_{(-c)(t)}},\bm{y_{(1:N)(1:t)}},\bm{v}) \\
& \times p(\bm{S_{(c)t}}|\bm{S_{(-c)(0:t+1)}},\bm{y_{(1:N)(1:t)}},\bm{v}).
\end{align*} Now, $P(\bm{S_{(c)t}}=\bm{s_{(c)t}}|\bm{S_{(-c)(0:t+1)}},\bm{y_{(1:N)(1:t)}},\bm{v})$ for $t=0,...,T$ and $\bm{s_{(c)t}} \in \{0,1\}^{n_c}$ {\color{black}(i.e. the presence/absence status of all locations in the block)} are known as the filtered probabilities. These are calculated using the forward part of the bFFBS algorithm, starting with $t=0$, we have,
\begin{align*}
P(\bm{S_{(c)0}}=\bm{s_{(c)0}}|\bm{S_{(-c)(0:1)}},\bm{v}) \propto \prod_{j \in (-c) }p(S_{j1}|\bm{S_{(c)0}}=\bm{s_{(c)0}},\bm{S_{(-c)0}},\bm{v}) \prod_{i \in c} P(S_{i0}=s_{i0}(\bm{s_{(c)0}})),
\end{align*} where $s_{i0}(\bm{s_{(c)0}})$ is the state indicator for location $i$ in $\bm{s_{(c)0}}$. Now for $t=1,...,T$ the predictive probabilities are first calculated and then used to calculate the filtered probabilities. The predictive probability is given by,
\begin{align}
\begin{split}
&P(\bm{S_{(c)t}}=\bm{s_{(c)t}}|\bm{S_{(-c)(0:t)}},\bm{y_{(1:N)(1:t-1)}},\bm{v}) = \\[5pt] 
& \,\,\, \sum_{\bm{s_{(c)t-1}}' \in \{0,1\}^{n_c}} P(\bm{S_{(c)t}}=\bm{s_{(c)t}}|\bm{S_{(c)(t-1)}}=\bm{s_{(c)t-1}}',\bm{S_{(-c)(t-1)}},\bm{y_{(1:N)(1:t-1)}},\bm{v}) \\ & \qquad  \qquad \qquad \times P(\bm{S_{(c)(t-1)}}=\bm{s_{(c)t-1}}'|\bm{S_{(-c)(0:t)}},\bm{y_{(1:N)(1:t-1)}},\bm{v}), \label{pred_prob}
\end{split}
\end{align} which is most efficiently calculated by multiplying the $2^{n_c} \times 2^{n_c}$ conditional transition matrix of $\bm{S_{(c)t}}$, $\Gamma(\bm{S_{(c)t}}|\bm{S_{(-c)(t-1)}},\bm{y_{(1:N)(1:t-1)}})$, transposed, by the vector of previous filtered probabilities. Note that an element of $\Gamma(\bm{S_{(c)t}}|\bm{S_{(-c)(t-1)}},\bm{y_{(1:N)(1:t-1)}})$ is given by,
\begin{align*}
&P(\bm{S_{(c)t}}=\bm{s_{(c)t}}|\bm{S_{(c)(t-1)}}=\bm{s_{(c)t-1}}',\bm{S_{(-c)(t-1)}},\bm{y_{(1:N)(1:t-1)}},\bm{v})= \\
&\prod_{i \in c} P(S_{it}=s_{it}(\bm{s_{(c)t}})|\bm{S_{(c)(t-1)}}=\bm{s_{(c)t-1}}',\bm{S_{(-c)(t-1)}},\bm{y_{(1:N)(1:t-1)}},\bm{v}),
\end{align*} where $s_{it}(\bm{s_{(c)t}})$ is the state indicator for location $i$ in $\bm{s_{(c)t}}$, which can be calculated from the individual area conditional transition matrices defined in equation (1) of the main text. The predictive probabilities are then used to calculate the filtered probabilities,
\begin{align}
\begin{split}
&P(\bm{S_{(c)t}}=\bm{s_{(c)t}}|\bm{S_{(-c)(0:t+1)}},\bm{y_{(1:N)(1:t)}},\bm{v}) \propto \\
& \,\,\,P(\bm{S_{(c)t}}=\bm{s_{(c)t}}|\bm{S_{(-c)(0:t)}},\bm{y_{(1:N)(1:t-1)}},\bm{v}) \prod_{i \in c}p(y_{it}|S_{it}=s_{it}(\bm{s_{(c)t}}),\bm{y_{(1:N)(1:t-1)}},\bm{v}) \\ 
& \times \prod_{j \in (-c) }p(S_{j(t+1)}|\bm{S_{(c)t}}=\bm{s_{(c)t}},\bm{S_{(-c)t}},\bm{y_{(1:N)(1:t)}},\bm{v}), 
\end{split}
\end{align} which involves the predictive probability. In (3), if $y_{it}>0$ and $s_{it}(\bm{s_{(c)t}})=0$, then $p(y_{it}|S_{it}=s_{it}(\bm{s_{(c)t}}),\bm{y_{(1:N)(1:t-1)}},\bm{v})=0$. This means the entire filtered probability can be set to 0 and, therefore, these conditions should be checked first to avoid unnecessary calculations. In the extreme case where $y_{it}>0$ for all $i \in c$, then $t$ can be skipped altogether as the entire vector of filtered probabilities is then known. Also, if $y_{it}>0$ but  $s_{it}(\bm{s_{(c)t}})=1$ then $p(y_{it}|S_{it}=s_{it}(\bm{s_{(c)t}}),\bm{y_{(1:N)(1:t-1)}},\bm{v})$ can be set to 1, since it will cancel in the calculation when normalizing. For $t=T$, there is no $\prod_{j \in (-c) }p(S_{j(t+1)}|\bm{S_{(c)t}}=\bm{s_{(c)t}},\bm{S_{(-c)t}},\bm{y_{(1:N)(1:t)}},\bm{v})$ term included.

After all filtered probabilities have been calculated then the backward sampling step is performed. Starting at $T=t$, $\bm{S_{(c)T}^{[m]}}$ is sampled categorically from the final filtered probabilities. Then, for $t=T-1,...,0$ we sample  $\bm{S_{(c)t}^{[m]}}$ from,
\begin{align}
\begin{split}
&P(\bm{S_{(c)t}^{[m]}}=\bm{s_{(c)t}}|\bm{S_{(c)(t+1)}^{[m]}},\bm{S_{(-c)(0:t+1)}},\bm{y_{(1:N)(1:t)},v)} \propto \\[5pt]
& \,\,\, p(\bm{S_{(c)(t+1)}^{[m]}}|\bm{S_{(c)t}^{[m]}}=\bm{s_{(c)t}},\bm{S_{(-c)t}},\bm{y_{(1:N)(1:t)}},\bm{v}) \\ & \times P(\bm{S_{(c)t}^{[m]}}=\bm{s_{(c)t}}|\bm{S_{(-c)(0:t+1)}},\bm{y_{(1:N)(1:t)}},\bm{v}). \label{backward_sampling}
\end{split}
\end{align} The first probability in (\ref{backward_sampling}) comes from $\Gamma(\bm{S_{(c)(t+1)}}|\bm{S_{(-c)t}},\bm{y_{(1:N)(1:t)}})$, the second probability is the filtered probability. Note that when $y_{it}>0$ the algorithm will always sample $S_{it}^{[m]}=1$ and, therefore, these can just be set ahead of running the MCMC as mentioned.

\subsection{Validating the algorithm}

To validate the bFFBS algorithm, we compared the posteriors produced by the bFFBS2 sampler, described in the main text, with those produced by the binary sampler. These were compared on the dengue data, whose fitting is described in the main text. The binary sampler is validated in the simulation study in Section 2 of this appendix. Therefore, the bFFBS2 sampler should produce the same posterior distributions as the binary sampler. 

\begin{figure}[t]
 	\centering
 	\includegraphics[width=.65\textwidth]{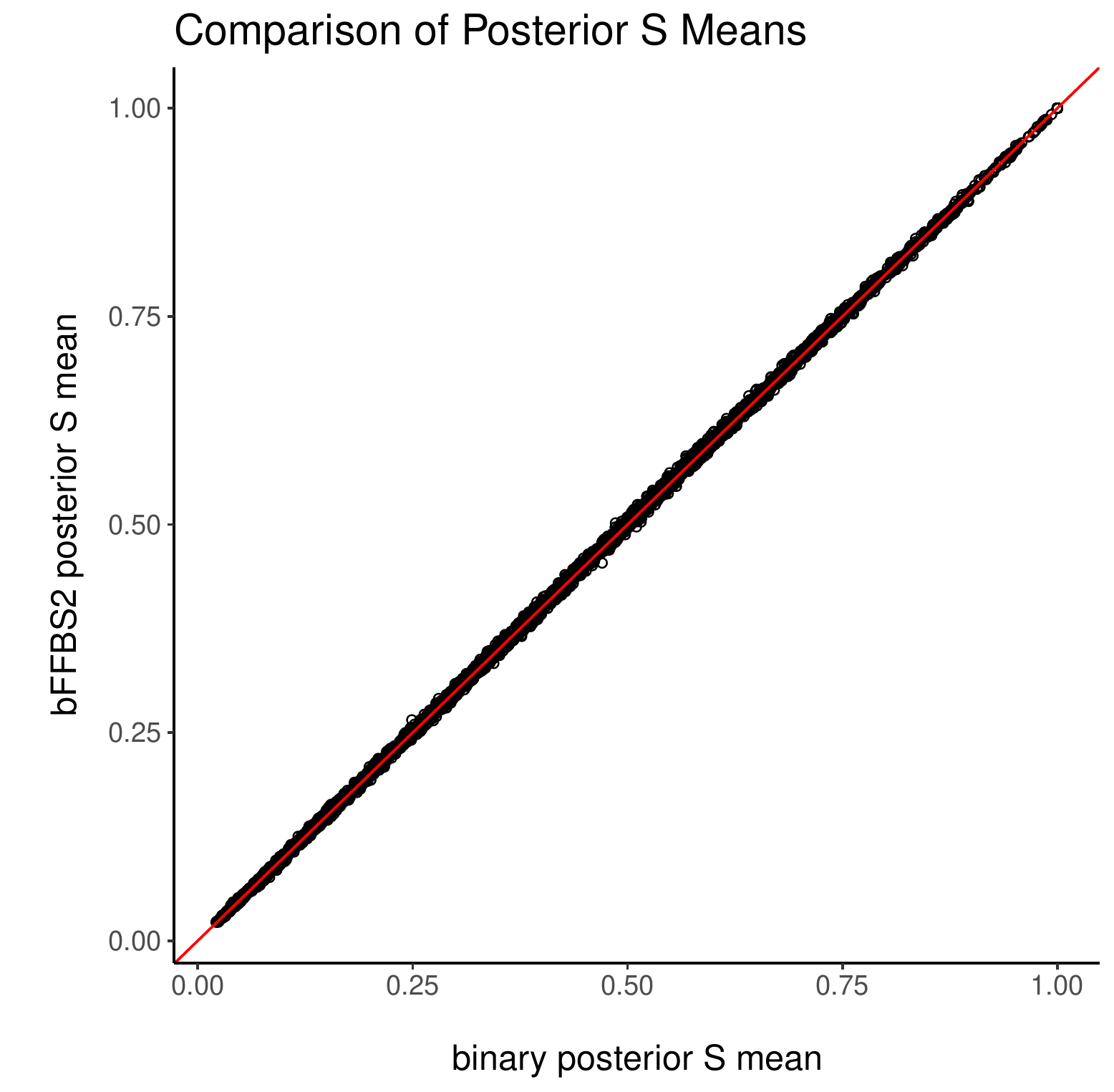}
	\caption{\label{sampler_compare} Comparison of the posterior means of $\bm{S}$ produced by the binary and bFFBS2 samplers.}
\end{figure}

In Figure 1, we show a plot comparing the posterior means of $\bm{S}$ from the two samplers. Both samplers produce the same posterior means for $\bm{S}$ within reasonable Monte Carlo error. Additionally, we compared the posterior means and 95\% posterior credible intervals for all elements of $\bm{v}$ (not shown). There was no meaningful difference in the posteriors of $\bm{v}$ produced by the two samplers.

\section{Simulation study}

We designed a simulation study to ensure the proposed hybrid Gibbs sampler can recover the true parameters of the {\color{black}ZS-CMSNB} model. {\color{black}In particular, we are interested in whether we can estimate heterogeneous effects of disease spread between areas and separate covariate effects for the reemergence and persistence of the disease. To simulate heterogeneous effects of between area disease spread, we assume there is a 30 percent chance that any two neighboring areas are separated by some barrier to disease spread that reduces the effect of disease spread between the two areas by 60 percent. This corresponds very similarly to the situation in \cite{smithPredictingSpatialDynamics2002} who looked at whether rivers reduce the effect of rabies spread between two neighboring areas. To simulate separate covariate effects we assume that temperature has half the effect on the persistence. We generated data from a ZS-CMSNB model with the following specifications for $\lambda_{it}$, $r_{it}$, $\bm{z_{it}}$, $\bm{z_{01,ijt}^{(c)}}$ and $\bm{z_{11,ijt}^{(c)}}$ (see equations (1)-(6) of the main text),
\begin{align}
\begin{split}
&\log(\lambda_{it}) = \beta_0 + \beta_1 \text{Temp}_{t-1}+ \beta_2 \text{HDI}_{i}, \\
& r_{it} = r \\
& \bm{z_{it}} = \left[\text{Temp}_{t-1}\right]^{T} \\
& \bm{z_{01,ijt}^{(c)}} = \left[\text{barr}_{ij}\right]^{T} \\
& \bm{z_{11,ijt}^{(c)}} = \emptyset,
\end{split} \label{sim_spec}
\end{align} where $\bm{v}=\left[\beta_0,\beta_1,\beta_2,r,\zeta_0,\zeta_1,\zeta_0^{(c)},\zeta_1^{(c)},\eta_0,\eta_1,\eta_0^{(c)}\right]^T \\ =\left[1,.1,10,1.5,-1.5,.1,.25,-.15,1.5,.05,.1\right]^T$, $\text{Temp}_{t-1}$ and $\text{HDI}_i$ are taken from the motivating example in the main text, and $\text{barr}_{ij}=1$ if locations $i$ and $j$ are separated by a barrier, for $i=1,...,160$ and $t=1,...,84$ like the motivating example. The neighborhood structure is that of Rio like the motivating example of the main text. Additionally we assumed $p(S_{i0}) \sim Bern(.5)$ for $i=1,...,N$. The simulations are set up so that the effects of temperature and between area disease spread on the reemergence and persistence of the disease are somewhat similar to our motivating example. Additionally, 50 percent of the counts are 0 and 50 percent of the zeroes come from the Markov chain, which represents a good amount of missing information.} We fit the {\color{black}ZS-CMSNB model (correctly specified)} to 100 replications of (\ref{sim_spec}). We ran {\color{black}80,000} iterations, with a burn-in of {\color{black}30,000}, of the Gibbs sampler on 3 chains started randomly from different points. For each replication, convergence of each parameter was checked using the effective sample size ($>$1000) and the Gelman-Rubin statistic ($<$1.05) \citep{plummerCODAConvergenceDiagnosis2006}. {\color{black}Additionally, as many zeroes often lead to model instability and identifiability issues in complex zero inflated models \citep{agarwalZeroinflatedModelsApplication2002}, we also ran a set of simulations at 80 percent zeroes by changing $\beta_0$ from 1 to -1.} 

\begin{figure}[t]
 	\centering
 	\includegraphics[width=.95\textwidth]{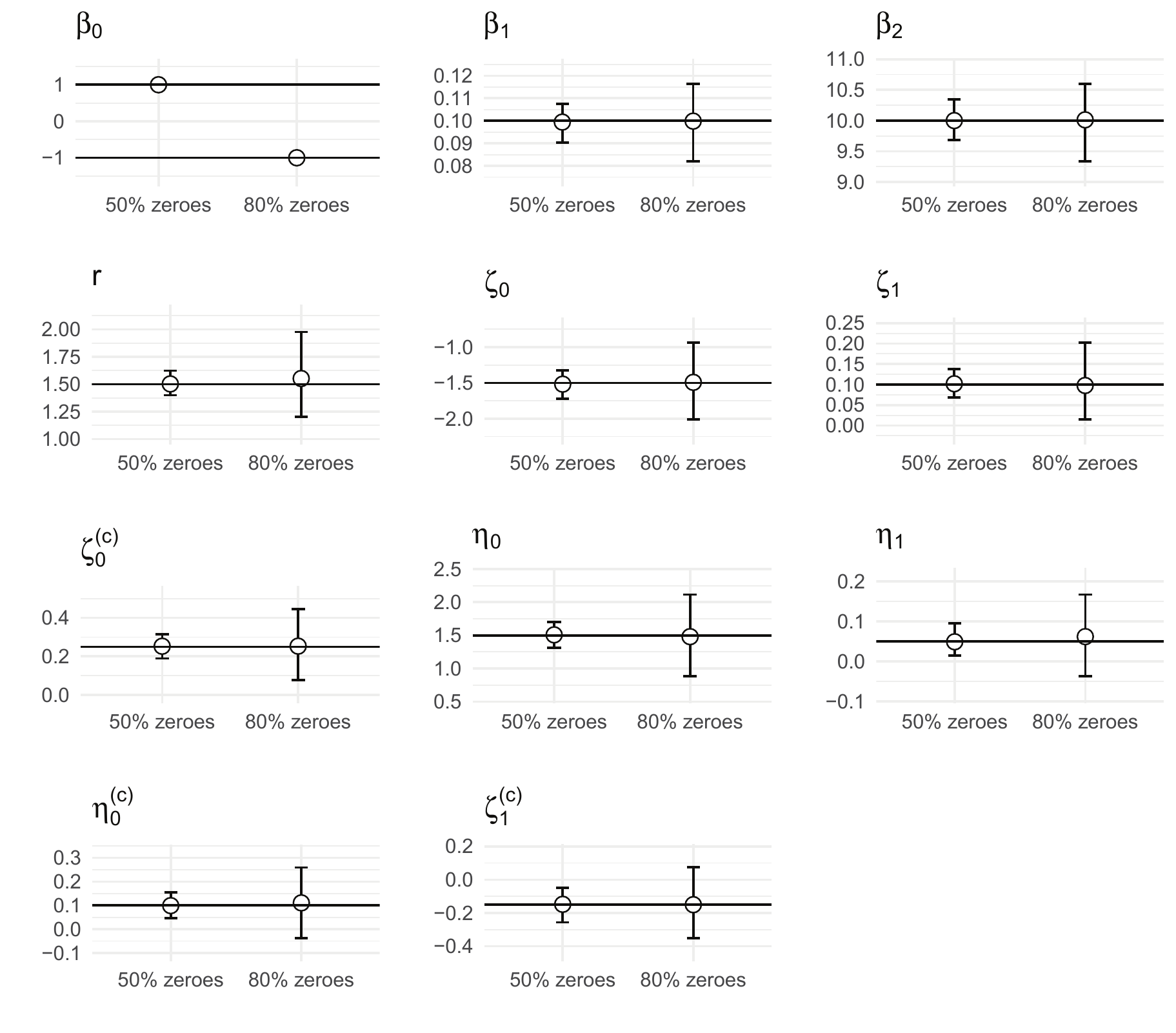}
	\caption{Averages and 95\% quantiles for the sampling distributions (100 replications) of the posterior means {\color{black}from a simulation study with 50 percent and 80 percent 0s. Horizontal lines drawn at true parameter values.}} \label{ZS-CMSP_sim}
\end{figure} 

{\color{black}The sampling distributions of the posterior means from each set of simulations can be seen in Figure \ref{ZS-CMSP_sim}. Figure \ref{ZS-CMSP_sim} shows that while the precision of the posterior means at 80 percent zeroes is much less then at 50 percent zeroes, the posterior means still have minimum bias in either scenario. Additionally, the average coverage of the 95\% posterior credible intervals at 80 percent zeroes was 94.3\% (min=92.7\%, max=95.3\%) and at 50 percent zeroes was 94.4\% (min=91.4\%, max=95.9\%). However, we did notice some convergence issues at 80 percent 0s that were not present at 50 percent 0s. We had to run the model for 5 times as long at 80 percent zeroes and around 5 percent of the simulations did not converge, while all simulations converged at 50 percent zeroes. 

In conclusion, our Gibbs sampler applied to the ZS-CMSNB model is able to estimate heterogeneous between area effects of disease spread and separate covariate effects with minimum bias and good coverage. However, at a large number of 0s (80 percent) the precision of the estimates are quite low and there can be rare convergence issues.}

\section{Monte Carlo Approximations for the Posterior Predictive Distributions}

We are interested in the K step ahead posterior predictive distribution of both disease presence, $p(S_{i(T+K)}|\bm{y})$, and the cases counts, $p(y_{i(T+K)}|\bm{y})$, for $K=1,2,...$ and $i=1,...,N$. We assume only dependence on the previous times counts for simplicity in notation, then we have,
\begin{align}
\begin{split}
p(S_{i(T+K)}|\bm{y}) =\int &p(S_{i(T+K)}|\bm{S_{T+K-1}},\bm{y_{T+K-1}},\bm{\theta}) \\
&\times p(\bm{y_{T+K-1}}|\bm{S_{T+K-1}},\bm{y_{T+K-2}},\bm{\beta})p(\bm{S_{T+K-1}}|\bm{S_{T+K-2}},\bm{y_{T+K-2}},\bm{\theta}) \\ 
& \dotsc \times p(\bm{y_{T+1}}|\bm{S_{T+1}},\bm{y_{T}},\bm{\beta})p(\bm{S_{T+1}}|\bm{S_{T}},\bm{y_{T}},\bm{\theta}) \\
& \times p(\bm{v}|\bm{y}) \,d\bm{y_{T+K-1}}d\bm{S_{T+K-1}} \dotsc d\bm{y_{T+1}}d\bm{S_{T+1}}d\bm{\beta}d\bm{\theta}, \label{pres_K_step}
\end{split}
\end{align} and,
\begin{align}
\begin{split}
p(y_{i(T+K)}|\bm{y}) &=\int p(y_{i(T+K)}|S_{i(T+K)},\bm{y_{T+K-1}},\bm{\beta})p(S_{i(T+K)}|\bm{S_{T+K-1}},\bm{y_{T+K-1}},\bm{\theta}) \\
& \qquad \times p(\bm{y_{T+K-1}}|\bm{S_{T+K-1}},\bm{y_{T+K-2}},\bm{\beta})p(\bm{S_{T+K-1}}|\bm{S_{T+K-2}},\bm{y_{T+K-2}},\bm{\theta}) \\ 
& \qquad  \dotsc \times p(\bm{y_{T+1}}|\bm{S_{T+1}},\bm{y_{T}},\bm{\beta})p(\bm{S_{T+1}}|\bm{S_{T}},\bm{y_{T}},\bm{\theta}) \\
& \qquad  \times p(\bm{v}|\bm{y}) \, d S_{i(T+K)}d\bm{y_{T+K-1}} d\bm{S_{T+K-1}} \dotsc d\bm{y_{T+1}}d\bm{S_{T+1}}d\bm{\beta}d\bm{\theta} \\[5pt]
& = \int \left[ p(y_{i(T+K)}|S_{i(T+K)}=1,\bm{y_{T+K-1}},\bm{\beta})P(S_{i(T+K)}=1|\bm{S_{T+K-1}},\bm{y_{T+K-1}},\bm{\theta}) \right. \\[5pt]
&  \left. \qquad +I\left[y_{i(T+K)}=0\right]\left(1-P(S_{i(T+K)}=1|\bm{S_{T+K-1}},\bm{y_{T+K-1}},\bm{\theta}\right) \right] \\ 
& \qquad \times p(\bm{y_{T+K-1}}|\bm{S_{T+K-1}},\bm{y_{T+K-2}},\bm{\beta})p(\bm{S_{T+K-1}}|\bm{S_{T+K-2}},\bm{y_{T+K-2}},\bm{\theta}) \\ 
& \qquad  \dotsc \times p(\bm{y_{T+1}}|\bm{S_{T+1}},\bm{y_{T}},\bm{\beta})p(\bm{S_{T+1}}|\bm{S_{T}},\bm{y_{T}},\bm{\theta}) \\
& \qquad  \times p(\bm{v}|\bm{y}) \,d\bm{y_{T+K-1}} d\bm{S_{T+K-1}} \dotsc d\bm{y_{T+1}}d\bm{S_{T+1}}d\bm{\beta}d\bm{\theta}. \label{int_K_step}
\end{split}
\end{align} The above integrals can be approximated through Monte Carlo integration, 
\begin{align}
\begin{split}
p(S_{i(T+K)}|\bm{y}) & \approx \frac{1}{Q-M} \sum_{m=M+1}^{Q} p(S_{i(T+K)}|\bm{S_{T+K-1}^{[m]}},\bm{y_{T+K-1}^{[m]}},\bm{\theta^{[m]}}), \label{approx_K_step_pres}
\end{split}
\end{align} and, \small
\begin{align}
\begin{split}
p(y_{i(T+K)}|\bm{y}) & \approx \frac{1}{Q-M} \sum_{m=M+1}^{Q} \left[ p(y_{i(T+K)}|S_{i(T+K)}=1,\bm{y_{T+K-1}}^{[m]},\bm{\beta}^{[m]})P(S_{i(T+K)}=1|\bm{S_{T+K-1}}^{[m]},\bm{y_{T+K-1}}^{[m]},\bm{\theta}^{[m]}) \right. \\[5pt]
&  \left. \qquad +I\left[y_{i(T+K)}=0\right]\left(1-P(S_{i(T+K)}=1|\bm{S_{T+K-1}}^{[m]},\bm{y_{T+K-1}}^{[m]},\bm{\theta}^{[m]}\right) \right], \label{approx_K_step}
\end{split}
\end{align} \normalsize where the superscript $[m]$ denotes a draw from the posterior distribution of the variable ($\bm{y_t^{[m]}}=\bm{y_t}$ if $t \leq T$), $M$ is the size of the burn-in sample, $Q$ is the total MCMC sample size and $I[\cdot]$ is an indicator function. 

Substituting $K=1$ into (\ref{approx_K_step_pres}) approximates the one step ahead posterior predictive distribution of disease presence in location $i$,
\begin{align}
\begin{split}
P(S_{i(T+1)}=1|\bm{y}) & \approx \frac{1}{Q-M} \sum_{m=M+1}^{Q} P(S_{i(T+1)}=1|S_{iT}^{[m]},\bm{S_{(-i)T}^{[m]}},\bm{y_{T}},\bm{\theta^{[m]}}), \label{approx_one_step_risk}
\end{split}
\end{align} for $i=1,...,N$, where the superscript $[m]$ denotes a draw from the posterior distribution of the parameter, $M$ is the size of the burn-in sample and $Q$ is the total MCMC sample size. We can similarly approximate the one step ahead posterior predictive distribution of the reported cases in location $i$, 
\begin{align}
\begin{split}
p(y_{i(T+1)}|\bm{y}) & \approx  \frac{1}{Q-M} \sum_{m=M+1}^{Q} \left[ p(y_{i(T+1)}|S_{i(T+1)}=1,\bm{y_{T}},\bm{\beta}^{[m]})P(S_{i(T+1)}=1|S_{iT}^{[m]},\bm{S_{(-i)T}^{[m]}},\bm{y_{T}},\bm{\theta^{[m]}}) \right. \\[5pt]
&  \left. +I\left[y_{i(T+1)}=0\right]\left(1-P(S_{i(T+1)}=1|S_{iT}^{[m]},\bm{S_{(-i)T}^{[m]}},\bm{y_{T}},\bm{\theta^{[m]}})\right) \right], \label{approx_one_step}
\end{split}
\end{align} for $i=1,...,N$, where $I[\cdot]$ is an indicator function. Therefore, the one step ahead posterior predictive distribution of the counts is zero inflated, where the mixing probability depends on past histories of the states and counts. We can expand (\ref{approx_one_step_risk}) to gain a better understanding of this one step ahead prediction for the risk of disease presence in location $i$,
\begin{align}
\begin{split}
&P(S_{i(T+1)}=1|S_{iT}^{[m]},\bm{S_{(-i)(T)}^{[m]}},\bm{y_{T}},\bm{\theta^{[m]}})= p01_{i(T+1)}^{[m]}(1-S_{iT}^{[m]})+p11_{i(T+1)}^{[m]}S_{iT}^{[m]},\text{ where,} \\[5pt]
&\text{logit}(p01_{i(T+1)}^{[m]})= \zeta_0^{[m]} + \bm{z_{i(T+1)}^T} \bm{\zeta^{[m]}} + \sum_{j \in NE(i)} {\color{black}\phi_{01,j \to i}^{T \to (T+1) \, [m]}} \, S_{jT}^{[m]},\\[5pt]
&\text{logit}(p11_{i(T+1)}^{[m]}) = \eta_0^{[m]} +  \bm{z_{i(T+1)}^T} \bm{\eta^{[m]}}+\sum_{j \in NE(i)} {\color{black}\phi_{11,j \to i}^{T \to (T+1) \, [m]}} \, S_{jT}^{[m]}. \label{expand_one_step_risk}
\end{split}
\end{align} Even if for $j \in NE(i)$ $y_{jT}=0$, $S_{jT}^{[m]}$ could be 1 for many $m$ if there is a high chance the disease went undetected in neighbor $j$. Similarly, if $y_{iT}=0$, $S_{iT}^{[m]}$ could be 1 for many $m$ if there is a high chance the disease is undetected in location $i$. Therefore, this warning system accounts for the fact that the disease may already have been circulating undetected in location $i$ for some time. It also considers the possible risk of spread from neighboring locations with 0 reported cases that have a high chance of the disease being undetected. More traditional warning systems, based on autoregressive models \citep{chanDailyForecastDengue2015}, cannot account for these important risks. This makes our warning system particularly useful in cities like Rio where underreporting is a major issue, and the disease may circulate and spread unnoticed.

For arbitrary $K$ step ahead temporal predictions, we use a simulation procedure \citep{fruhwirth-schnatterFiniteMixtureMarkov2006} to draw samples from the posterior predictive distributions. Algorithm \ref{alg:alg1} will obtain realizations from the posterior predictive distribution of the cases, $y_{i(T+k)}^{[m]} \sim p(y_{i(T+k)}|\bm{y})$, and the presence of the disease, $S_{i(T+k)}^{[m]} \sim p(S_{i(T+k)}|\bm{y})$, for $i=1,...,N$, $k=1,...,K$ and $m=M+1,...,Q$. However, as $S_{i(T+k)}^{[m]}$ can only take two values, 0 or 1, it is difficult to interpret the uncertainty around this prediction for the presence of the disease. Therefore, instead of using summaries of $S_{i(T+k)}^{[m]}$ we use summaries of $P(S_{i(T+k)}=1|S_{i(T+k-1)}^{[m]},\bm{S_{(-i)(T+k-1)}^{[m]}},\bm{y_{T+k-1}^{[m]}},\bm{\theta^{[m]}})$.
\begin{algorithm}
\caption{Posterior Predictive Simulation}
\label{alg:alg1}
\SetAlgoLined
\For{$m$ in  $M+1:Q$}{
\For{$k$ in  $1:K$}{
\For{$i$ in $1:N$}{
1. Draw $S_{i(T+k)}^{[m]}$ from $p(S_{i(T+k)}|\bm{S_{T+k-1}^{[m]}},\bm{y_{T+k-1}^{[m]}},\bm{\theta^{[m]}})$, where $\bm{y_{T}^{[m]}}=\bm{y_{T}}$, \\[5pt]
\qquad $p(S_{i(T+k)}|\bm{S_{T+k-1}^{[m]}},\bm{y_{T+k-1}^{[m]}},\bm{\theta^{[m]}})$=Bern$\left(\pi_{i(T+k)}^{[m]}\right)$, where, \\[5pt]         \qquad $\pi_{i(T+k)}^{[m]}=p01_{i(T+k)}^{[m]}(1-S_{i(T+k-1)}^{[m]})+p11_{i(T+k)}^{[m]}S_{i(T+k-1)}^{[m]}$. \\[10pt]
2. Draw $y_{i(T+k)}^{[m]}$ from $p(y_{i(T+k)}|S_{i(T+k)}^{[m]},\bm{y_{T+k-1}^{[m]}},\bm{\beta^{[m]}})$, where $\bm{y_{T}^{[m]}}=\bm{y_{T}}$, \\[5pt]
\qquad $p(y_{i(T+k)}|S_{i(T+k)}^{[m]},\bm{y_{T+k-1}^{[m]}},\bm{\beta^{[m]}})={\color{black}NB(S_{i(T+k)}^{[m]} \lambda_{i(T+k)}^{[m]},r_{i(T+k)}^{[m]})}$.
}}}
\end{algorithm}

\begin{figure}[!htb]
 	\centering
 	\includegraphics[width=.95\textwidth]{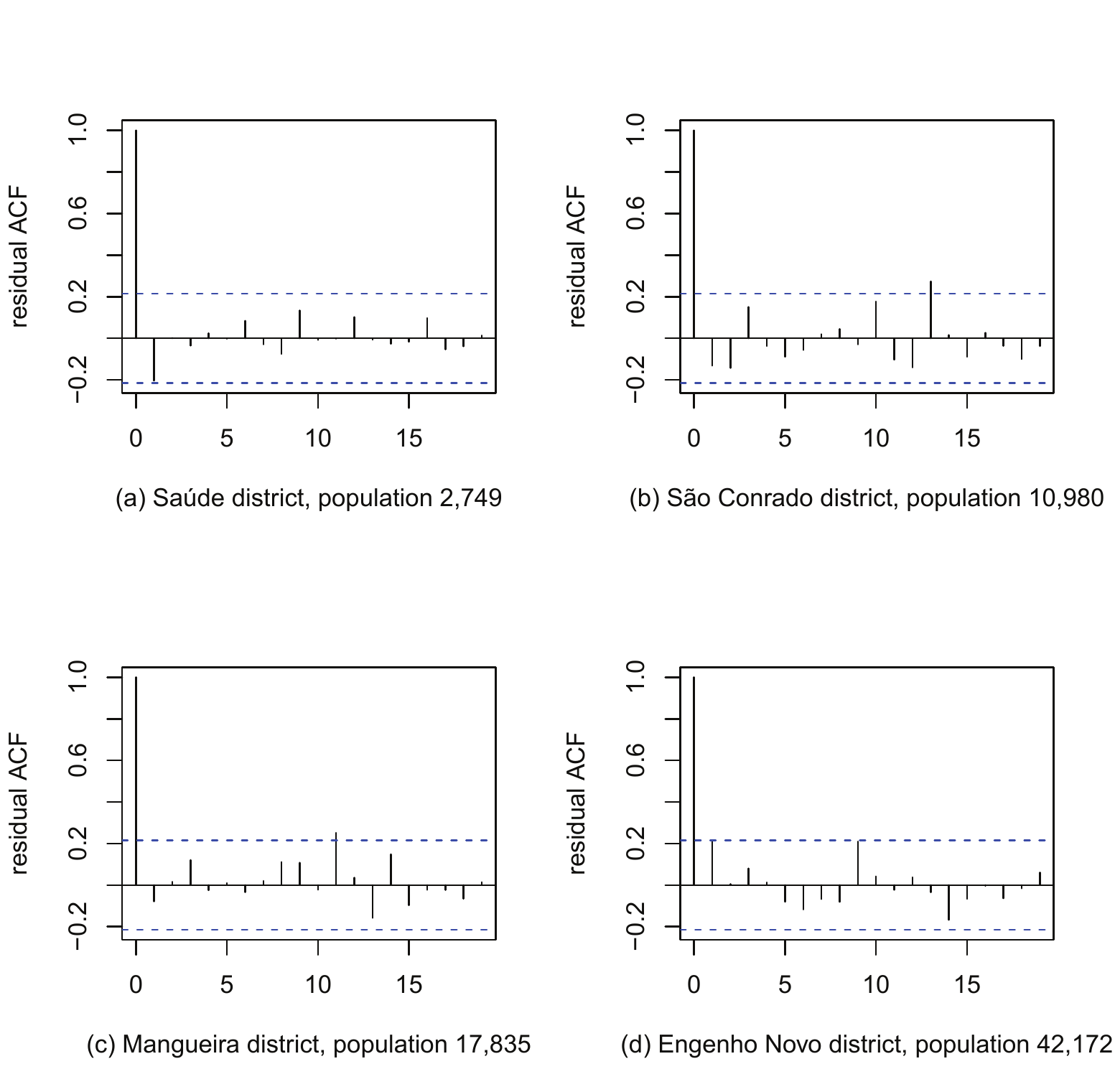}
	\caption{{\color{black}ACF of the Pearson residuals for the 4 districts from Figure 6 of the main text.}} \label{PRESID}
\end{figure}

{\color{black}
\section{Pearson Residuals}

A common model diagnostic in time series analysis is the autocorrelation function (ACF) of the Pearson residuals \citep{bracherEndemicepidemicModelsDiscretetime2020a}. Following Section 3.2, we define the Pearson residual in area $i$ for our model as,
\begin{align*}
&\text{Pres}_{it} = \frac{y_{it}-E[y_{it}|\bm{y^{(t-1)}},\bm{S_{(-c)(0:t)}},\bm{v}]}{\sqrt{Var[y_{it}|\bm{y^{(t-1)}},\bm{S_{(-c)(0:t)}},\bm{v}]}}.
\end{align*} We estimate $E[y_{it}|\bm{y^{(t-1)}},\bm{S_{(-c)(0:t)}},\bm{v}]$ and $Var[y_{it}|\bm{y^{(t-1)}},\bm{S_{(-c)(0:t)}},\bm{v}]$ using the sample mean and variance of $y_{it}^{c1*[m]}$. Figure \ref{PRESID} (below) shows the ACF of the Pearson residuals for the 4 districts from Figure 6 of the main text. There are not many significant autocorrelations. Overall, the figure suggests we are capturing well the structure present in the data.}

\newpage 

\section*{Acknowledgements}
This work is part of the PhD thesis of D. Douwes-Schultz under the supervision of A. M. Schmidt in the Graduate Program of Biostatistics at McGill University, Canada. Schmidt is grateful for financial support from the Natural Sciences and Engineering Research Council (NSERC) of Canada (Discovery Grant RGPIN-2017-04999). {\color{black}Douwes-Schultz is grateful for financial support from IVADO and the Canada First Research Excellence Fund / Apogée (PhD Excellence Scholarship). This research was enabled in part by support provided by Calcul Québec (www.calculquebec.ca) and Compute Canada (www.computecanada.ca).}

\bibliography{supplement}